\documentclass[10pt,twoside]{article}
\usepackage{Latex-document}
\usepackage{amssymb,amsfonts}
\usepackage{amsmath}
\usepackage{epic,graphics}
\usepackage{amsthm}
\usepackage{amsxtra}

\newcommand{\ad}{\mathop{\rm ad}}
\newcommand{\codim}{\mathop{\rm codim}}
\newcommand{\sdim}{\mathop{\rm sdim}}
\newcommand{\Der}{\mathop{\rm Der \, }}
\newcommand{\diag}{{\rm diag}}
\renewcommand{\div}{\mathop{\rm div}}
\newcommand{\End}{\mathop{\rm  End \, }}
\newcommand{\even}{\mathop{\rm  even \, }}
\renewcommand{\Im}{\mathop{\rm  Im \, }}
\newcommand{\Ind}{\mathop{\rm  Ind \, }}
\newcommand{\mult}{\mathop{\rm mult}}
\newcommand{\odd}{\mathop{\rm  odd \, }}
\newcommand{\Res}{\mathop{\rm Res}}
\newcommand{\tr}{{\rm tr}}
\newcommand{\Vect}{\mathop{\rm Vect}}
\newcommand{\Vir}{\mathop{\rm Vir}}

\newcommand{\fa}{{\mathfrak a}}
\newcommand{\fg}{{\mathfrak g}}
\newcommand{\fh}{{\mathfrak h}}
\newcommand{\fn}{{\mathfrak n}}
\newcommand{\fp}{{\mathfrak p}}

\newcommand{\A}{\mathcal{A}}
\renewcommand{\H}{\mathcal{H}}
\renewcommand{\O}{\mathcal{O}}

\newcommand{\CC}{{\mathbb C}}
\newcommand{\QQ}{{\mathbb Q}}
\newcommand{\ZZ}{{\mathbb Z}}

\newcommand{\st}[1]{\ensuremath{^{\scriptstyle \textrm{#1}}}}

\renewcommand{\bar}{\overline}

\def\examenum{%
  \ifnum \@enumdepth >\thr@@\@toodeep\else
    \advance\@enumdepth\@ne
    \edef\@enumctr{enum\romannumeral\the\@enumdepth}%
      \expandafter
      \list
        \csname label\@enumctr\endcsname
        {\usecounter\@enumctr\def\makelabel##1{\hss\llap{##1}}}%
  \fi}
\let\examenum =\endlist

\makeatletter

\newcommand{\arabicparenlist}{
  \renewcommand{\theenumi}{\arabic{enumi}}%
  \renewcommand{\labelenumi}{(\theenumi)}%
}

\newcommand{\romanparenlist}{
  \renewcommand{\theenumi}{\roman{enumi}}%
  \renewcommand{\labelenumi}{(\theenumi)}%
}

\newcommand{\alphaparenlistii}{
  \renewcommand{\theenumii}{\alph{enumii}}%
  \renewcommand{\labelenumii}{(\theenumii)}%
}

\newcounter{bean}
\newenvironment{deflist}[1]%
    {
      \begin{list}{\bf #1\arabic{bean}}
         {\usecounter{bean}
              \setcounter{bean}{-1}
          \labelsep=1em
          \settowidth{\labelwidth}{#1\thebean:}
          \addtolength{\labelwidth}{1.1ex}
          \leftmargin=\labelwidth
          \addtolength{\leftmargin}{\labelsep} }
          \renewcommand{\makelabel}[1]{\upshape(##1)\hfil}
    }    {\end{list}}

\makeatletter
\renewcommand\section{\@startsection {section}{1}{\z@}%
                                   {-3.5ex \@plus -1ex \@minus -.2ex}%
                                   {2.3ex \@plus.2ex}%
                                   {\normalfont\large\bfseries}}
\renewcommand\subsection{\@startsection{subsection}{2}{\z@}%
                                     {-3.25ex\@plus -1ex \@minus -.2ex}%
                                     {0ex \@plus .0ex}%
                                     {\normalfont\normalsize}}

\newtheorem{theorem}{Theorem}
\newtheorem*{theorem*}{Theorem}

 \makeatletter
   \def\examenum{%
     \ifnum \@enumdepth >\thr@@\@toodeep\else
       \advance\@enumdepth\@ne
       \edef\@enumctr{enum\romannumeral\the\@enumdepth}%
         \list
           {\csname label\@enumctr\endcsname}%
           {\usecounter\@enumctr
             \addtolength{\leftmargin}{-\leftmargin}
             \settowidth{\labelwidth}{(99)}
             \itemindent = \labelwidth
             \addtolength{\itemindent}{\labelsep}
             \def\makelabel##1{{##1}\hfill}
             }%
     \fi}
   
   \makeatother

 \theoremstyle{definition}
\newtheorem*{examples*}{Examples}

\theoremstyle{remark}
\newtheorem*{remark*}{Remark}

\def\volumeno{I}
\title{\bf Classification of Supersymmetries\vskip 6mm}
\author{Victor G. Kac\thanks{Department of Mathematics, MIT,
Cambridge, MA 02139, USA. E-mail:
kac@math.mit.edu}\vspace*{-0.5cm}}
\date{\vspace{-8mm}}

\begin{document}

\maketitle

\thispagestyle{first} \setcounter{page}{319}

\begin{abstract}

\vskip 3mm

In the first part of my talk I will explain a solution to the
extension of Lie's problem on classification of "local continuous
transformation groups of a finite-dimensional manifold" to the
case of supermanifolds. (More precisely, the problem is to
classify simple linearly compact Lie superalgebras, i.e.
toplogical Lie superalgebras whose underlying space is a
topological product of finite-dimensional vector spaces). In the
second part I will explain how this result is used in a
classification of superconformal algebras. The list consists of
affine superalgebras and certain super extensions of the Virasoro
algebra. In the third part I will discuss representation theory of
affine superalgebras and its relation to "almost" modular forms.
Furthermore, I will explain how the quantum reduction of these
representations leads to a unified representation theory of super
extensions of the Virasoro algebra. In the forth part I will
discuss representation theory of exceptional simple
infinite-dimensional linearly compact Lie superalgebras and will
speculate on its relation to the Standard Model.

\vskip 4.5mm


\end{abstract}

\vskip 12mm

\section*{Introduction}

\vskip-5mm \hspace{5mm}

The theory of Lie groups and Lie algebras began with the 1880
paper \cite{L} of S.~Lie where he posed the problem of
classification of ``local continuous transformation groups of a
finite-dimensional manifold'' $M$ and gave a solution to this
problem when $\dim M = 1$ and $2$.

The most important part of Lie's problem is the classification of
the corresponding Lie algebras of vector fields on $M$ up to
``formal'' isomorphism.  A more invariant (independent of $M$)
formulation is to classify linearly compact Lie algebras,
i.e.,~topological Lie algebras whose underlying space is a
topological product of discretely topologized finite-dimensional
vector spaces \cite{GS}, \cite{G2}.  (Of course, it is well-known
  that it is impossible to classify even all finite-dimensional
  Lie algebras.  What is usually meant be a ``classification'' is
  a  complete list of simple algebras (no non-trivial ideas) and
  a description of semisimple algebras (no abelian ideals) in
  terms of simple ones.)

It turned out that a solution to this problem requires quite
different methods in the cases of finite-dimensional and
infinite-dimensional groups.  The most important advance in the
finite-dimensional case was made by W.~Killing and E.~Cartan at
the end of the 19\st{th} century who gave the celebrated
classification of simple finite-dimensional Lie algebras over
$\CC$.  The infinite-dimensional case was studied by Cartan in a
series of papers written in the beginning of the 20\st{th}
century, which culminated in his classification of
infinite-dimensional ``primitive' Lie algebras of vector fields on
a finite-dimensional manifold \cite{C}.

The advent of supersymmetry in theoretical physics in the 1970s
motivated work on the ``super'' extension of Lie's problem.  In
the finite-dimensional case the latter problem was settled in
\cite{K2}.  However, it took another 20~years before the problem
was solved in the infinite-dimensional case \cite{K7}, \cite{CK2},
\cite{CK3}.

In the first part of my talk I will explain the classification of
simple linearly compact Lie superalgebras.  Remarkably, unlike in
the Lie algebra case, the approach, based on the ideas of the
papers \cite{GS}, \cite{W}, \cite{K1} and \cite{G2}, is very
similar in the finite- and infinite-dimensional cases.

The advent of conformal field theory in the mid-1980s motivated
the work on classification and representation theory of
superconformal algebras.  In the second part of my talk I will
explain how the classification of infinite-dimensional simple
linearly compact Lie algebras is applied to classification of
``linear'' simple superconformal algebras.  A complete list
consists of the affine superalgebras, and of several series and
one exceptional example of super extensions of the Virasoro
algebra \cite{FK}.  (The most famous of these super extensions is
the $N=2$ superconformal algebra, which plays a fundamental role
in the mirror symmetry theory.)

In the third part of my talk I will discuss representation theory
of affine superalgebras \cite{KW3}, \cite{KW4}.  The key property
of ``admissible'' representations of affine algebras is that their
characters are modular functions.  This is not so in the super
case---for some mysterious reason, modular functions get replaced
by closely related but more general functions, like Appell's
function \cite{KW4}.

Next, I will explain how the quantum reduction of ``admissible''
representations of affine superalgebras leads to a unified
representation theory of (not necessarily linear) super extensions
of the Virasoro algebra \cite{KRW},\cite{KW5}.  This gives rise to
a large class of supersymmetric rational conformal field theories.

In the last part of my talk I will discuss representation theory
of exceptional infinite-dimensional simple linearly compact Lie
superalgebras \cite{KR1}--\cite{KR4}.  I am convinced that this
theory may have applications to ``real'' physics.   The main
reason for this belief is the exceptional Lie superalgebra
$E(3|6)$ whose maximal compact group of automorphisms is the gauge
group of the Standard Model ($=$ a quotient of $SU_3 \times SU_2
\times U_1$ by a cyclic group of order $6$).  Furthermore,
representation theory of $E(3|6)$ accurately predicts the number
of generations of leptons ($=3$), but not so accurately the number
of generations of quarks ($=5$) \cite{KR2}.  It is also striking
that the inclusion of the gauge group of the Standard Model in
$SU_5$, which is the gauge group of the Grand Unified Model,
extends to the inclusion of $E(3|6)$ in $E(5 | 10)$, the largest
exceptional linearly compact Lie superalgebra.

\section{Classification of simple linearly compact Lie superalgebras.}
\label{sec:1}

\vskip-5mm \hspace{5mm}

\subsection{}
\label{sec:1.1} First, recall some basic superalgebra terminology.  A
\emph{superalgebra} is simply a $\ZZ / 2 \ZZ$-graded algebra:
\begin{displaymath}
  S = S_{\bar{0}} + S_{\bar{1}} \, , \hbox{ where }
  S_{\alpha} S_{\beta} \subset S_{\alpha + \beta} \, ,
  \alpha, \beta \in \ZZ /2 \ZZ  = \{ \bar{0},\bar{1} \} \, .
\end{displaymath}
If $a \in S_{\alpha}$, one says that the \emph{parity} $p(a)$ is
equal to $\alpha$. A \emph{derivation $D$} of parity $p(D)$ of a
superalgebra $S$ is a vector space endomorphism satisfying
condition
\begin{displaymath}
  D(ab) = (Da)b+ (-1)^{p(D)p(a)} a(Db) \, .
\end{displaymath}
The sum $\Der S$ of the spaces of derivations of parity $\bar{0}$
and $\bar{1}$ is closed under the super bracket:
\begin{displaymath}
  [D, D_1] = DD_1 - (-1)^{p(D)p(D_1)} D_1 D \, .
\end{displaymath}
This super bracket satisfies super analogs of anticommutativity
and Jacobi identity, hence defines what is called a \emph{Lie
  superalgebra}.  (The super anticommutativity axiom is
$[a,b]=-(-1)^{p(a)p(b)} [b,a]$, and the super Jacobi identity
axiom means that the operator $(\ad a) b:= [a,b]$ is a
derivation.)

One of the basic constructions is the \emph{superization} which
basically amounts to adding anticommuting indeterminates.  In
other words, given an algebra (associative or Lie) $\A$ we
consider the Grassmann algebra $\A\langle n \rangle$ in $n$
anticommuting indeterminates $\xi_1 , \ldots , \xi_n$ over $\A$.
This algebra carries a canonical $\ZZ / 2 \ZZ$-gradation  defined
by letting
  $p(\A)=\overline{0}, \quad p (\xi_i) = \overline{1} $.
   If $\O_m$ denotes the algebra of
formal power series over $\CC$ in $m$ indeterminates, then
$\O_m\langle n \rangle$ is the algebra over $\CC$ of formal power
series in $m$ commuting indeterminates $x=(x_1, \ldots , x_m)$ and
$n$ anticommuting indeterminates $\xi = (\xi_1 , \ldots , \xi_n)$:
$  x_ix_j=x_jx_i, \quad x_i\xi_j = \xi_j x_i, \quad
  \xi_i \xi_j =-\xi_j \xi_i$.

Note that the associative superalgebra $\O_m \langle n \rangle$ is
linearly compact with respect to the topology for which the powers
of the augmentation ideal $(x_1,\ldots ,x_m \, , \,
  \xi_1 , \ldots , \xi_n)$ form a fundamental system of
neighborhoods of $0$.  The algebra $\A \langle n \rangle$ has odd
(i.e.,~of parity $\bar{1}$) derivations $\partial / \partial
\xi_i$ defined by
\begin{displaymath}
  \frac{\partial}{\partial \xi_i} (a) =0 \hbox{ for }
  a \in \A, \quad \frac{\partial }{\partial \xi_i}
  (\xi_j) = \delta_{ij} \, ,
\end{displaymath}
and these derivations anticommute, i.e.,~$[\partial /\partial
\xi_i \, , \, \partial /\partial \xi_j] =0$.

The first basic example of a linearly compact Lie superalgebra is
the Lie superalgebra denoted by $W (m | n)$, of all continuous
derivations of the topological superalgebra $\O_m \langle n
\rangle$:
\begin{displaymath}
  W(m|n) = \left\{ \sum^m_{i=1} P_i (x,\xi)
    \frac{\partial}{\partial x_i} + \sum^n_{j=1}
    Q_j (x,\xi) \frac{\partial}{\partial \xi_j} \right\} \, ,
\end{displaymath}
where $P_i (x,\xi), Q_j (x,\xi) \in \O_m\langle n \rangle$.  In a
more geometric language, this is the Lie superalgebra of all
formal vector fields on a supermanifold of dimension $(m|n)$.

\subsection{}
\label{sec:1.2}

Cartan's theorem \cite{C} states that a complete list of
infinite-dimensional linearly compact simple Lie algebras over
$\CC$ consists of four series:  the Lie algebra $W_m (= W(m | 0))$
of all formal vector fields on an $m$-dimensional manifold, and
its subalgebras $S_m$ of divergenceless vector fields $(m>1)$,
$H_m$ of Hamiltonian vector fields ($m$ even), $K_m$ of contact
vector fields ($m$ odd).

There is a unique way to extend \emph{divergence} from $W_m$ to
$W(m|n)$ such that the divergenceless vector fields form a
subalgebra:
\begin{displaymath}
  \div \left(\sum_i P_i \frac{\partial}{\partial x_i} + \sum_j
    Q_j \frac{\partial}{\partial \xi_j} \right)
  = \sum_i \frac{\partial P_i}{\partial x_i} + \sum_j (-1)^{p(Q_j)}
  \frac{\partial Q_j}{\partial \xi_j} \, ,
\end{displaymath}
and the super analog of $S_m$ is
\begin{displaymath}
  S(m|n) = \{ X \in W(m|n) |  \div X =0 \} \, .
\end{displaymath}

In order to define super analogs of the Hamiltonian and contact
Lie algebras $H_m$ and $K_m$, introduce a super analog of the
algebra of differential forms \cite{K2}.  This is an associative
superalgebra over $\O_m\langle n \rangle$, denoted by
$\Omega(m|n)$, on generators $dx_1 , \ldots , dx_m$,
$d\xi_1,\ldots,d\xi_n$ and defining relations: $  dx_idx_j
=-dx_jdx_i, \quad dx_i d\xi_j=d\xi_j dx_i, \quad
  d\xi_i d\xi_j = d\xi_j d\xi_i$,
and the $\ZZ /2\ZZ$-gradation defined by: $
p(x_i)=p(d\xi_j)=\overline{0}, \quad
  p(\xi_j)=p(dx_i)=\overline{1}$.
This superalgebra is linearly compact in the topology defined by
powers of the augmentation ideal. The topological superalgebra
$\Omega(m|n)$ carries a unique continuous derivation $d$ of parity
$\overline{1}$ such that $  d(x_i)=dx_i, \quad d(\xi_j)=d\xi_j,
\quad d(dx_i)=0,\quad
  d(d\xi_j)=0$.
The operator $d$ has all the usual properties, e.g.: $  df=\sum_i
dx_i \frac{\partial f}{\partial x_i} +
  \sum_j \frac{\partial f}{\partial \xi_j} \, d\xi_j
  \hbox{ for }f \in \O_m\langle n \rangle, \hbox{ and } d^2=0$.
As usual, for any $X \in W(m|n)$ one defines a derivation
$\iota_X$ (contraction along $X$) of the superalgebra
$\Omega(m|n)$ by the properties (here $x$ stands for $x$ and
$\xi$): $  p(\iota_X) = p(X) + \overline{1}, \quad
  \iota_X (x_j)=0, \quad
  \iota_X (dx_j) =(-1)^{p(X)} X(x_j)$.
The action of any $X \in W(m|n)$ on $\O_m\langle n \rangle$
extends in a unique way to the action by a derivation of
$\Omega(m|n)$ such that $[X,d]=0$.  This is called Lie's
derivative and is usually denoted by $L_X$, but we shall write $X$
in place of $L_X$ unless confusion may arise.  One has the usual
Cartan's formula for this action:  $L_X = [d,\iota_X]$.

Using this action, one can define super-analogs of the Hamiltonian
and contact Lie algebras for any $n \in \ZZ_+$:
\begin{eqnarray*}
 H(m|n) &=& \{ X \in W(m|n) |  X\omega_s =0 \} \, , \\
%
\noalign{\hbox{\hbox{where } $ m=2k   \hbox{  and  } \omega_s =
\sum^k_{i=1} dx_i \wedge dx_{k+i}
      + \sum^n_{j=1} (d \xi_j)^2$,}}\\[-2ex]
%
  K(m|n) &=& \{ X \in W(m|n)|  X \omega_c = f\omega_c \} \, , \\
%
 \noalign{\hbox{\hbox{where }$   m=2k+1\, , \, \omega_c = dx_m + \sum^k_{i=1} x_i dx_{k+i}
      + \sum^n_{j=1} \xi_j \, d\xi_j ,  \hbox{ and }
      f\in  \O_m\langle n \rangle$.}}
\end{eqnarray*}

Note that $W(0|n)$, $S(0|n)$ and $H(0|n)$ are finite-dimensional
Lie superalgebras.  The Lie superalgebras $W(0|n)$ and $S(0|n)$
are simple iff $n \geq 2$ and $n \geq 3$, respectively.  However,
$H(0|n)$ is not simple as its derived algebra $H'(0|n)$ has
codimension~$1$ in $H(0|n)$, but $H'(0|n)$ is simple iff $n\geq
4$.  Thus, in the Lie superalgebra case the lists of simple
finite- and infinite-dimensional algebras are much closer related
than in the Lie algebra case.

These four series of Lie superalgebras are infinite-dimensional if
$m \geq 1$, in which case they are simple except for $S(1|n)$. The
derived algebra $S'(1|n)$ has codimension~$1$ in $S(1|n)$, and
$S'(1|n)$ is simple iff $n \geq 2$.

Remarkably it turned out that the above four series  do not
exhaust all infinite-dimensional simple linearly compact Lie
superalgebras (as has been suggested in \cite{K2}). Far from it!

As was pointed out by several mathematicians, the Schouten bracket
\cite{SV} makes the space of polyvector fields on a
$m$-dimensional manifold into a Lie superalgebra.  The formal
analog of this is the following fifth series of superalgebras,
called by physicists the Batalin-Vilkoviski algebra ($H$ stands
here for ``Hamiltonian'' and $O$ for ``odd''):
\begin{displaymath}
  HO(m|m) =\{ X \in W(m|m) |   X\omega_{os} =0 \} \, ,
\end{displaymath}
where $\omega_{os} = \sum^m_{i=1} \, dx_i d\xi_i$ is an ``odd''
symplectic form.  Furthermore, unlike in the $H(m|n)$ case, not
all vector fields of $HO(m|n)$ have zero divergence, which gives
rise to the sixth series:
\begin{displaymath}
  SHO(m|m) =\{ X \in HO(m|m)|  \div X=0 \} \, .
\end{displaymath}
The seventh series is the odd analog of $K(m|n)$ \cite{ALS}:
\begin{displaymath}
KO(m|m+1) =\{ X \in W(m|m+1) |  X\omega_{oc} =f\omega_{oc} \} \, ,
\end{displaymath}
where $\omega_{oc}=d\xi_{m+1} + \sum^m_{i=1} (\xi_i \, dx_i + x_i
\, d \xi_i)$ is an odd contact form.  One can take again the
divergence $0$ vector fields in $KO(m|m+1)$ in order to construct
the eighth series, but the situation is more interesting. It turns
out that for each $\beta \in \CC$ one can define the deformed
divergence $\div_{\beta}X$ \cite{Ko}, \cite{K7}, so that $\div =
\div_0$ and
\begin{displaymath}
  SKO(m|m+1;\beta) =\{ X \in KO(m|m+1) |   \hbox{div}_{\beta} X=0 \}
\end{displaymath}
is a subalgebra.  The superalgebras $HO(m|m)$ and $KO(m|m+1)$ are
simple iff $m \geq 2$ and $m \geq 1$, respectively.  The derived
algebra $SHO'(m|m)$ has codimension~$1$ in $SHO(m|m)$, and it is
simple iff $m \geq 3$.  The derived algebra $SKO'(m|m+1;\beta)$ is
simple iff $m \geq 2$, and it coincides with $SKO(m|m+1;\beta)$
unless $\beta =1$ or $\frac{m-2}{m}$ when it has codimension~$1$.

Some of the examples described above have simple ``filtered
deformations'', all of which can be obtained by the following
simple construction.  Let $L$ be a subalgebra of $W(m|n)$, where
$n$ is even.  Then it happens in three cases that
\begin{displaymath}
  L^{\sim}:=(1+\Pi^n_{j=1} \xi_j)L
\end{displaymath}
is different from $L$, but is closed under bracket.  As a result
we get the following three series of superalgebras:
$S^{\sim}(0|n)$ \cite{K2}, $SHO^{\sim}(m|m)$ \cite{CK2} and
$SKO^{\sim}(m|m+1;\frac{m+2}{m})$ \cite{Ko} (the constructions in
\cite{Ko} and \cite{CK2} were more complicated).  We thus get the
ninth and the tenth series of simple infinite-dimensional Lie
superalgebras:
\begin{eqnarray*}
  SHO^{\sim}(m|m), \quad m \geq 2 \,, \, m \even \, ,\\
SKO^{\sim}(m|m+1;1+2/m) \, ,\, m \geq 3, m \odd \, .
\end{eqnarray*}

A surprising discovery was made in \cite{Sh1} where the existence
of three exceptional simple infinite-dimensional Lie superalgebras
was announced.  The proof of the existence along with one more
exceptional example was given in \cite{Sh2}.  An explicit
construction of these four examples was given later in \cite{CK3}.
The fifth exceptional example was found in the work on conformal
algebras \cite{CK1} and independently in \cite{Sh2}.  (The alleged
sixth exceptional example $E(2|2)$ of \cite{K7} turned out to be
isomorphic to $SKO(2|3;1)$ \cite{CK3}.)

Now I can state the first main theorem.

\begin{theorem} {\rm \cite{K7}}
  \label{th:1}
  The complete list of simple infinite-dimensional linearly
  compact Lie superalgebras consists of ten series of examples
  described above and five exceptional examples:  $E(1|6)$,
  $E(3|6)$, $E(3|8)$, $E(4|4)$, and $E(5|10)$.
\end{theorem}

Here and before the notation $X(m | n)$ means that this
superalgebra can be embedded in $W (m | n)$ and that this
embedding is minimal possible; $E$ stands for ``exceptional''.

\begin{remark*}
  The local classification of transitive primitive (i.e.,~leaving
  no invariant fibrations) actions on a (super)manifold $M$ is
  equivalent to the classification of all ``primitive'' pairs
  $(L,L_0)$, where $L$ is a linearly compact Lie (super)algebra
  and $L_0$ is a maximal open subalgebra (such that $\dim L/L_0 =
  \dim M$) without non-zero ideals of $L$. If $L$ is simple,
  choosing any maximal open subalgebra $L_0$, we get a primitive
  pair $(L,L_0)$.  One can show that if, in addition, $L$ is a
  Lie algebra and $\dim L=\infty$, there exists a unique such
  $L_0$.  According to Cartan's theorem, the remaining
  infinite-dimensional primitive pairs are the Lie algebras
  obtained from $S_n$ and $H_n$ by adding the Euler operator $E$.
  Using the structure results on general transitive linearly
  compact Lie algebras \cite{G1}, it is not difficult to reduce
  the classification of infinite-dimensional primitive pairs to
  the classification of simple infinite-dimensional linearly
  compact Lie algebras (cf.~\cite{G2}).  Such a reduction is
  possible also in the Lie superalgebra case, but it is much more
  complicated for two reasons: (a)~a simple linearly compact Lie
  superalgebra may have several maximal open subalgebras (see
  \cite{CK3} for a classification),
  (b)~construction of arbitrary primitive pairs in terms of
  simple primitive pairs is more complicated in the superalgebra
  case (see~\cite{K8}).

\end{remark*}

\subsection{}
\label{sec:1.3}

Here I will describe the classification of finite-dimensional simple Lie
superalgebras.  We already have four ``non-classical'' series:  $W (0 |
n)$, $S(0 | n)$, $H' (0 | n)$ and $S^{\sim} (0 |n)$.  The four
``classical'' series are constructed as follows. Introduce the following
even and odd Euler operators $E= \sum_i x_i \frac{\partial}{\partial
x_i} +
  \sum_j \xi_j \frac{\partial}{\partial \xi_j} \in W (m
|n)$ and $ E_o = \sum_i x_i \frac{\partial}{\partial \xi_i} +
  \sum_i \xi_i \frac{\partial}{\partial x_i}\in W(m |m)$.
Let
\begin{eqnarray*}
  s \ell (m | n) &=&
      \{ X \in S (m|n)| [E,X] =0 \} \, , \\
 spo  (m | n) &=&
      \{ X \in H (m|n)| [E,X] =0 \} \, , \\
 p  (m | m) &=&
      \{ X \in SHO (m|m)| [E,X] =0 \} \, , \\
 q  (m | m) &=&
      \{ X \in W (m|m)| [E_o,X] =0 \} \, .
\end{eqnarray*}
The Lie algebras $s\ell_m = s\ell (m|0)$, $sp_m =spo (m|0)$, $so_n
= spo (0|n)$ are simple.  Furthermore, $s\ell (m|n)$ are simple
for $m \neq n$, all $spo (m|n)$ and $p(m|m)$ $(m \geq 3)$ are
simple. The superalgebra $s\ell (m|m)$ contains $1$-dimensional
ideal $\CC E$ and $s\ell (m|m)/\CC E$ is simple for $m \geq 2$.
Finally, the derived algebra $q' (m|m)$ has codimension $1$ in
$q(m|m)$ and $q'(m|m)/\CC E$ is simple for $m \geq 3$.

\begin{theorem}{\rm \cite{K2}}
  \label{th:2}
The complete list of simple finite-dimensional Lie superalgebras
consists of eight series of examples described above, the
exceptional Lie superalgebras $F(4)$ and $G(3)$ of dimension~$40$
and $31$, respectively, a $1$-parameter family of $17$-dimensional
exceptional Lie superalgebras $D (2,1;a)$, and the five
exceptional Lie algebras.

\end{theorem}

\subsection{Plan of the proof of Theorem~\ref{th:1}.} \quad
\label{sec:1.4}

Step 1.~~Introduce Weisfeiler's filtration \cite{W} of a simple linearly
compact Lie superalgebra $L$.  For
  that choose a maximal open subalgebra $L_0$ of $L$ and a
  minimal subspace $L_{-1}$ satisfying the properties:
$    L_{-1} \supsetneqq L_0 , \quad [L_0,L_{-1}] \subset L_{-1}$.
 (Geometrically this corresponds to a choice of a primitive action
of $L$ and an invariant irreducible differential system.)  The
pair $L_{-1},L_0$ can be included in a unique filtration: $
L=L_{-d} \supsetneq L_{-d+1} \supset \cdots \supset
  L_{-1} \supset L_0 \supset L_1 \supset \cdots$,
called \emph{Weisfeiler's filtration} of \emph{depth}~$d$. (In the
Lie algebra case, $d>1$ only for $K_m$, when $d=2$, but in the Lie
superalgebra case, $d>1$ in the majority of cases.)  The
associated to Weisfeiler's filtration $\ZZ$-graded Lie
superalgebra is of the form $GrL = \Pi_{j \geq -d} \fg_j$, and has
the following properties:

\begin{deflist}{G}
  \item 
    $\dim \fg_j < \infty$ (since $\codim L_0 < \infty$),

  \item 
    $\fg_{-j} = \fg^j_{-1}$ for $ j \geq 1$ (by maximality of $L_0$),

  \item 
    $[x, \fg_{-1}]=0$ for $x \in \fg_j$, $j \geq 0 \Rightarrow
    x=0$ (by simplicity of $L$),

    \item 
       $\fg_0$-module $\fg_{-1}$ is irreducible (by choice of
       $L_{-1}$), and faithful (by (G2)).
\end{deflist}

Weisfeiler's idea was that property (G3) is so restrictive, that
it should lead to a complete classification of $\ZZ$-graded Lie
algebras satisfying (G0)--(G3).  (Incidentally, the
infinite-dimensionality of $L$ and hence of $Gr L$, since $L$ is
simple, is needed only in order to conclude that $\fg_1 \neq 0$.)
 This indeed turned out to be the case \cite{K1}.  In fact, my
idea was to replace the condition of finiteness of the depth by
finiteness of the growth, which allowed one to add to the
Lie-Cartan list some new Lie algebras, called nowadays affine
Kac-Moody algebras.

However, unlike in the Lie algebra case, it is impossible to
classify all finite-dimensional irreducible faithful
representations of Lie superalgebras.  One needed a new idea to
make this approach work.

Step 2.~~The main new idea is to choose $L_0$ to
  be invariant with respect to all inner automorphisms of $L$
  (meaning to contain all even $\ad$-exponentiable elements of
  $L$).  A non-trivial point is the existence of such $L_0$.
  This is proved by making use of the characteristic
  supervariety, which involves rather difficult arguments of
  Guillemin \cite{G2}.

Next, using a normalizer trick of Guillemin \cite{G2}, I prove,
for this choice of $L_0$, the following very powerful restriction
on the $\fg_0$-module $\fg_{-1}$ (at this point $\dim L=\infty$ is
used):

\begin{deflist}{G}
\setcounter{bean}{3}
   \item 
       $[\fg_0, x] =\fg_{-1}$ for any non-zero even element $x$ of
       $\fg_{-1}$.

\end{deflist}
%

Step 3.~~Consider a faithful irreducible
  representation of a Lie superalgebra $\fp$ in a
  finite-dimensional vector space $V$.  This representation is
  called \emph{strongly transitive} if
$    \fp \cdot x = V$ for any non-zero even element $x\in V$.  By
properties  (G0), (G3)  and (G4), the $\fg_0$-module $\fg_{-1}$ is
strongly transitive.

In order to demonstrate the power of this restriction, consider
first the case when $\fp$ is a Lie algebra and $V$ is purely even.
Then the strong transitivity simply means that $V \backslash
\{0\}$ is a single orbit of the Lie group $P$ corresponding to
$\fp$. It is rather easy to see that the only strongly transitive
subalgebras $\fp$ of $g \ell_{V}$ are $g\ell_{V}$, $s \ell_{V}$,
$sp_{V}$ and $csp_{V}$.  These four cases lead to $Gr L$, where
$L= W_n$, $S_n$, $H_n$ and $K_n$, respectively.

In the super case the situation is much more complicated.  First
we consider the case of ``inconsistent gradation'', meaning that
$\fg_{-1}$ contains a non-zero even element.  The classification
of such strongly transitive modules is rather long and the answer
consists of a dozen series and a half dozen exceptions (see
\cite{K7}, Theorem~3.1).  Using similar restrictions on $\fg_{-2},
\fg_{-3}, \ldots $, we obtain a complete list of possibilities for
$  GrL_{\leq}:= \oplus_{j \leq 0} \fg_j$, in the case when
$\fg_{-1}$ contains non-zero even elements.  It turns out that all
but one exception are not exceptions at all, but correspond to the
beginning members of some series.  As a result, only $E(4|4)$
``survives''.

Step 4.~~Next, we turn to the case of a consistent
  gradation, i.e.,~when $\fg_{-1}$ is purely odd.  But then
  $\fg_0$ is an ``honest'' Lie algebra, having a faithful
  irreducible representation in $\fg_{-1}$ (condition (G4)
  becomes vacuous).  An explicit description of such
  representations is given by the classical Cartan-Jacobson
  theorem.  In this case I use the ``growth'' method developed in
  \cite{K1}  and \cite{K2} to determine a complete list of
  possibilities for $Gr L_{\leq}$.  This case produces mainly the
  (remaining four) exceptions.

Step 5~is rather long and tedious \cite{CK3}  .
  For each $GrL_{\leq}$ obtained in Steps~3 and~4 we determine
  all possible ``prolongations'', i.e.,~infinite-dimensional
  $\ZZ$-graded Lie superalgebras satisfying (G2), whose negative part
  is the given $Gr L_{\leq}$.

Step 6.~~It remains to reconstruct $L$ from $GrL$,
  i.e.,~to find all possible filtered simple linearly compact Lie
  superalgebras $L$ with given $Gr L$ (such an $L$ is called a
  simple filtered deformation of $GrL$).  Of course,
  there is a trivial filtered deformation:  $GrL :=
  \Pi_{j \geq -d} \fg_j$, which is simple iff $GrL$ is.
It is proved in \cite{CK2} by a long and tedious calculation that
only $SHO(m|m)$ for $m \even \geq 2$ and $SKO(m|m+1;
  \frac{m+2}{m})$ for $m \odd \geq 3$ have a non-trivial simple
filtered deformation, which are the ninth and tenth series.  It
would be nice to have a more conceptual proof. Recall that
$SH0(m|m)$ is not simple, though it does have a simple filtered
deformation.  Note also that in the Lie algebra case all filtered
deformations are trivial.

\subsection{Plan of the proof of Theorem~\ref{th:2}.}\quad
\label{sec:1.5}

The key idea is the same as in the proof of Theorem~\ref{th:1}. Choose a
maximal subalgebra $L_0$ of a simple finite-dimensional Lie superalgebra
$L$ containing the even part of $L$. It is easy to see then that $L_{-1}
=L$, so that the corresponding Weisfeiler's filtration has depth~$1$.
Hence $Gr L$
 has the form:
$   Gr L = \oplus^N_{j=-1} \fg_j$. This gradation is consistent
and, of course, satisfies conditions (G0)--(G3).  There are two
cases.

Case 1.~~$N \geq 1$.  Then we apply the growth method (as in
Step~4 of Sec.~\ref{sec:1.4}) to obtain a complete list of
possibilities for $GrL_{\leq}$.  Then, as in Step~5 of
Sec.~\ref{sec:1.4} we determine all prolongations of each $Gr
L_{\leq}$ (all of them will be subalgebras of $W(0,\dim \fg_1)$),
and all filtered deformations of these prolongations.  This case
produces all ``non-classical'' series, and also $s\ell (m|n)$,
$spo (m|2)$ and $p(m|m)$.

Case 2.~~$N=0$.  Then  $L_{\bar{0}}$ is a semisimple Lie algebra
and its representation in $L_{\bar{1}}$ is irreducible.  The
Killing form on $L$ is either non-degenerate, in which case we
apply the standard Killing-Cartan techniques, or it is identically
zero.  In the latter case one uses Dynkin's index to find all
possibilities for the $L_{\bar{0}}$-module $L_{\bar{1}}$.

\subsection{}
\label{sec:1.6}

In order to describe the construction of the exceptional
infinite-dimensional Lie superalgebras (given in \cite{CK3}), I need to
make some remarks.  Let $\Omega_m = \Omega(m|0)$ be the algebra of
differential forms over $\O_m$, let $\Omega^k_{m}$ denote the space of
forms of degree $k$, and $\Omega^k_{m,c\ell}$ the subspace of closed
forms.  For any $\lambda \in \CC$ the representation of $W_m$ on
$\Omega^k_m$ can be ``twisted'' by letting
\begin{displaymath}
  X \mapsto L_X + \lambda \div X , \quad X \in W_m \, ,
\end{displaymath}
to get a new $W_m$-module, denoted by $\Omega^k_m(\lambda)$ (the same
can be done for $W(m|n)$).  Obviously, $\Omega^k_m(\lambda)=\Omega^k_m$
when restricted to $S_m$.  Then we have the following obvious
$W_m$-module isomorphisms: $\Omega^0_m \simeq \Omega^m_m (-1)$ and
$\Omega^0_m(1) \simeq \Omega^m_m$.  Furthermore, the map $X \mapsto
\iota_X (dx_1 \wedge \ldots \wedge dx_m)$ gives the following
$W_m$-module and $S_m$-module isomorphisms:
\begin{displaymath}
  W_m \simeq \Omega^{m-1}_m(-1), \quad S_m \simeq \Omega^{m-1}_{m,c\ell} \, .
\end{displaymath}
We shall identify the representation spaces via these
isomorphisms.

The simplest is the construction of the largest exceptional Lie
superalgebra $E(5|10)$.  Its even part is the Lie algebra $S_5$,
its odd part is the space of closed $2$-forms
$\Omega^2_{5,c\ell}$.  The remaining commutators are defined as
follows for $X \in S_5, \quad \omega,\omega' \in
\Omega^2_{5,c\ell}$:
\begin{displaymath}
  [X,\omega]=L_X\omega, \quad [\omega,\omega']=\omega \wedge \omega' \in \Omega^4_{5,c\ell}=S_5
  \, .
\end{displaymath}

Each quintuple of integers $(a_1,a_2,\ldots ,a_5)$ such that
$a=\sum_i a_i$ is even, defines a $\ZZ$-gradation of $E(5|10)$ by
letting:
\begin{displaymath}
  \deg x_i=- \frac{\partial}{\partial x_i} =a_i , \quad
  \deg dx_i =a_i -\tfrac{1}{4} a \, .
\end{displaymath}
The quintuple $(2,2, \ldots ,2)$ defines the (only) consistent
$\ZZ$-gradation, which has depth $2$:    $E(5|10) =\Pi_{j \geq
  -2} \fg_j$, and one has:
\begin{displaymath}
\fg_0 \simeq s \ell_5 \hbox{ and } \fg_{-1} \simeq \Lambda^2
\CC^5, \quad \fg_{-2} \simeq \CC^{5*} \hbox{ as $\fg_0$-modules.}
\end{displaymath}
Furthermore, $\Pi_{j \geq 0} \fg_j$ is a maximal open subalgebra
of $E(5|10)$ (the only one which is invariant with respect to all
automorphisms).  There are three other maximal open subalgebras in
$E(5|10)$, associated to $\ZZ$-gradations corresponding to
quintuples $(1,1,1,1,2)$, $(2,2,2,1,1)$ and $(3,3,2,2,2)$, and one
can show that these four are all, up to conjugacy, maximal open
subalgebras (cf.~\cite{CK3}).

Another important $\ZZ$-gradation of $E(5|10)$, which is, unlike
the previous four, by infinite-dimensional subspaces, corresponds
to the quintuple $(0,0,0,1,1)$ and has depth~$1$: $E(5|10)
=\Pi_{\lambda \geq -1} \fg^{\lambda}$.  One has: $\fg^0 \simeq
E(3|6)$ and the $\fg^{\lambda}$ form an important family of
irreducible $E(3|6)$-modules \cite{KR2}.  The consistent
$\ZZ$-gradation of $E(5|10)$ induces that of $\fg^0:E(3|6)=\Pi_{j
\geq -2} \fa_j$, where
\begin{displaymath}
  \fa_0 \simeq s\ell_3 \oplus s\ell_2 \oplus g\ell_1 , \quad
  \fa_{-1} \simeq\CC^3 \boxtimes \CC^2 \boxtimes \CC, \quad
  \fa_{-2}  \simeq \CC^3 \boxtimes \CC \boxtimes \CC \, .
\end{displaymath}
A more explicit construction of $E(3|6)$ is as follows \cite{CK3}:
the even part is $W_3 + \Omega^0_3 \otimes s\ell_2$, the odd part
is $\Omega^1_3 (-\tfrac{1}{2}) \otimes \CC^2$ with the obvious
action of the even part, and the bracket of two odd elements is
defined as follows:
\begin{displaymath}
  [\omega \otimes u,\omega' \otimes v] =
  (\omega \wedge \omega') \otimes (u \wedge v) +
  (d\omega \wedge \omega' + \omega \wedge d\omega') \otimes (u \cdot v) \, .
\end{displaymath}
Here the identifications $\Omega^2_3 (-1) =W_3$ and
$\Omega^0_3=\Omega^3_3 (-1)$ are used.

The gradation of $E(5|10)$ corresponding to the quintuple
$(0,1,1,1,1)$ has depth~$1$ and its $0$\st{th} component is
isomorphic to $E(1|6)$ (cf.~\cite{CK3}).

The construction of $E(4|4)$ is also very simple \cite{CK3}: The
even part is $W_4$, the odd part is $\Omega^1_4 (-\tfrac{1}{2})$
and the bracket of two odd elements is:
\begin{displaymath}
  [\omega,\omega'] = d\omega \wedge \omega'
+ \omega \wedge d\omega' \in \Omega^3_4(-1)=W_4 \, .
\end{displaymath}
The construction of $E(3|8)$ is slightly more complicated, and we
refer to \cite{CK3} for details.

\subsection{}
\label{sec:1.7}

All exceptional simple finite-dimensional Lie superalgebras (including
the exceptional Lie algebras) are obtained as special cases of the
following important construction \cite{K2}.  Let $I = \{ 1,\ldots ,r \}$
and let $I_{\bar{1}}$ be a subset of $I$, $I_{\bar{0}} = I \backslash
I_{\bar{1}}$.  Let $A = (a_{ij})_{i,j
  \in I}$ be a matrix over $\CC$.  We associate to the pair
$(A,I_{\bar{1}})$ a Lie superalgebra $\fg (A,I_{\bar{1}})$ as
follows.  Let $\tilde{\fg} (A,I_{\bar{1}})$ be the Lie
superalgebra on generators $e_i$, $f_i$, $h_i$ $(i \in I)$ of
parity $p(h_i)=\bar{0}$ for $i \in I$, $p(e_i) = p(f_i) = \alpha
\in \{ \bar{0} , \bar{1} \}$ for $i \in I_{\alpha}$, and the
following standard relations:
\begin{displaymath}
  \left[h_i , h_j\right] = 0 , \left[e_i , f_j \right] =
  \delta_{ij} h_i \, ,\,
  \left[h_i , e_j\right] = a_{ij} e_j , \left[h_i , f_j\right] = -a_{ij} f_j .
\end{displaymath}
Define a $\ZZ$-gradation $\tilde{\fg} (A,I_{\bar{1}}) =\oplus_{j
  \in \ZZ} \tilde{\fg}_j $ by letting $\deg h_i =0$, $\deg e_i =
-\deg f_i =1$.  Then $\tilde{\fg_0}$ is the $\CC$-span of $\{ h_i
\}_{i \in I}$, and we denote by $J (A,I_{\bar{1}})$ the sum of all
$\ZZ$-graded ideals of $\tilde{\fg} (A,I_{\bar{1}})$ that
intersect $\tilde{\fg}_0$ trivially.  We let

\begin{displaymath}
  \fg (A,I_{\bar{1}}) = \tilde{\fg} (A,I_{\bar{1}}) /
  J (A,I_{\bar{1}}) \, .
\end{displaymath}

Of course, if $A$ is the Cartan matrix of a simple
finite-dimensional Lie algebra $\fg$, then $\fg \simeq \fg
(A,\emptyset)$, the ideal $J (A,\emptyset)$ being generated by
``Serre relations''.  Likewise, generalized Cartan matrices give
rise to Kac-Moody Lie algebras \cite{K3}.

Consider the following matrices ($a \in \CC \backslash \{ 0 , -1
\}$):
\begin{eqnarray*}
  D_a =\left[
    \begin{array}{rrr}
      0 & -1 & -a \\
      -1 & 2 & 0 \\
      -1 & 0 & 2
    \end{array}\right] \, , \,
F= \left[
  \begin{array}{rrrr}
    0 & -1 & 0 & 0 \\
    -1 & 2 & -2 & 0\\
    0 & -1 & 2 & -1 \\
    0 & 0 & -1 & 2
  \end{array}\right] \, , \,
G = \left[
  \begin{array}{rrr}
    0 & -1 & 0\\
    -1 & 2 & -3 \\
    0 & -1 & 2
  \end{array} \right] \, .
\end{eqnarray*}
Then $D(2,1;\alpha) \simeq \fg (D_{\alpha}, \{ 1 \})$, $F(4)
\simeq \fg (F , \{ 1 \})$ and $G(3) \simeq \fg (G, \{ 1 \})$.
Note, however, that unlike in the Lie algebra case,
``inequivalent'' pairs $(A,I_{\bar{1}})$ may produce isomorphic
Lie superalgebras.  For example, in the cases $D(2,1;a)$, $F(4)$
and $G(3)$ there are $2$, $6$ and $4$ such pairs, respectively.

Finite-dimensional simple Lie superalgebras that are isomorphic to
$\fg (A, I_{\bar{1}})$ for some matrix $A$ are called \emph{basic}
(they will play an important role in the next parts of the talk).
The remaining basic simple Lie superalgebras (that are not Lie
algebras) are $s\ell (m|n)/\delta_{m,n} \CC E$ and $spo(m|n)$.

\section{A classification of superconformal algebras}
\label{sec:2}

\vskip-5mm \hspace{5mm}

Superconformal algebras have been playing an important role in
superstring theory and in conformal field theory.  Here I will
explain how to apply Theorem~\ref{th:1} to the classification of
``linear'' superconformal algebras.  By a (``linear'')
superconformal algebra I mean a Lie superalgebra $\fg$ spanned by
coefficients of a finite collection $F$ of fields such that the
following two properties hold:

\arabicparenlist

\begin{enumerate}
\item 
  for $a,b \in F$ the singular part of OPE is finite, i.e.,
  \begin{displaymath}
    [a(z),b(w)] = \sum_j c_j (w) \partial^j_w \delta (z-w) \quad
    \hbox{(a finite sum),}
\hbox{ where all } c_j (w) \in \CC [\partial_w]F \, ,
  \end{displaymath}
\vspace{-.25in}
\item 
  $\fg$ contains no non-trivial ideals spanned by coefficients of
  fields from a $\CC [\partial_w]$-submodule of
  $\CC [\partial_w]F$.
\end{enumerate}

(Recall that a field is a formal expression $a(z) = \sum_{n \in
  \ZZ} a_n z^n$, where $a_n \in \fg$ and $z$ is an indeterminate,
and $\delta (z-w) = z^{-1} \sum_{n \in \ZZ} (w/z)^n$ is the formal
$\delta$-function.  See \cite{K5} for details.)

This problem goes back to the physics paper \cite{RS}, some
progress in its solution was made in \cite{K6} and a complete
solution was given in \cite{FK}.  (A complete classification even
in the ``quadratic'' case seems to be a much harder problem,
see~\cite{FL} and Section~4 below for some very interesting
examples.)  The simplest example is the loop algebra $\tilde{\fg}=
\CC [x,x^{-1}] \otimes \fg$ (= centerless affine Kac-Moody
superalgebra), where $\fg$ is a simple finite-dimensional Lie
superalgebra.  Then $F=\{
  a(z) = \sum_{n \in \ZZ} (x^n \otimes a)z^{-n-1}\}_{a \in
    \fg}$, and $[a(z),b(w)]=[a,b](w) \delta (z-w)$.  The next
example is the Lie algebra $\Vect \CC^{\times}$ of regular vector
fields on $\CC^{\times}$ (= centerless Virasoro algebra); $F$
consists of one field, the Virasoro field $  L(z)=-\sum_{n \in
\ZZ} (x^n \frac{d}{dx})z^{-n-1}$, and $
         [L(z),L(w)]=$\break $\partial_w L(w) \delta (z-w)+2L(w) \delta'_w(z-w)$.

One of the main theorems of \cite{DK} states that these are all
examples in the Lie algebra case.  The strategy of the proof is
the following.  Let $\partial =\partial_z$ and consider the
(finitely generated) $\CC [\partial]$-module $R=\CC [\partial]F$.
Define the ``$\lambda$-bracket'' $R \otimes R \to \CC [{\lambda}]
\otimes R$ by the formula: $  [a_{\lambda}b]=\sum_j \lambda^j
c_j$. This satisfies the axioms of a \emph {conformal
(super)algebra} (see \cite{DK}, \cite{K5}), similar to the Lie
(super)algebra axioms:

\romanparenlist

\begin{enumerate}
\item 
  $[\partial a_{\lambda}b] =-\lambda [a_{\lambda}b]$, $[a_{\lambda}
  \partial b] = (\partial + \lambda) [a_{\lambda}b]$,

\item 
  $[a_{\lambda}b]=-(-1)^{p(a)P(b)}[b_{-\lambda-\partial}a]$,

\item 
  $[a_{\lambda}[b_{\mu}c]] = [[a_{\lambda}b]_{\lambda + \mu }c] +
  (-1)^{p(a)p(b)} [b_{\mu}[a_{\lambda}c]]$.
\end{enumerate}

The main observation of \cite{DK} is that a conformal
(super)algebra is completely determined by the Lie (super)algebra
spanned by all coefficients of negative powers of $z$ of the
fields $a(z)$ from $F$, called the \emph{annihilation algebra},
along with an even surjective derivation of the annihilation
algebra. Furthermore, apart from the case of current algebras, the
completed annihilation algebra turns out to be an
infinite-dimensional simple linearly compact Lie (super)algebra of
growth~$1$. Since in the Lie algebra case the only such example is
$W_1$, the proof is finished.

In the superalgebra case the situation is much more interesting
since there are many infinite-dimensional simple linearly compact
Lie superalgebras of growth $1$.  By Theorem~\ref{th:1}, the
complete list is as follows:
\begin{displaymath}
  W(1|N), \quad S'(1|N), \quad K(1|N) \,\, \hbox{ and } E(1|6)
  \, .
\end{displaymath}
In all cases, except the second, there is a unique, up to
conjugacy, even surjective derivation, hence a unique
corresponding superconformal algebra.  They are denoted by
$W_{(N)}$, $K_{(N)}$ if $N \neq 4$, $K'_{(4)}$ and $CK_{(6)}$,
respectively.  The Lie superalgebras $W_{(N)}$ and $K_{(N)}$ are
constructed in the same way as $W(1|N)$ and $K(1|N)$, except that
one replaces $\O_1 \langle N \rangle$ by $\CC [x,x^{-1}] \langle N
\rangle $.  The construction of the exceptional superconformal
algebra $CK_{(6)}$ is more difficult, and may be found in
\cite{CK1} or \cite{K6}.  However, $S'(1|N)$ has two families of
even surjective derivations.  The corresponding superconformal
algebras are derived algebras of
\begin{displaymath}
  S_{(N), \epsilon ,a} = \{ X \in W_{(N)} |
  \div (e^{ax} (1+\epsilon \xi_1 \ldots \xi_N)X)=0 \} \, ,
  \quad a \in \CC , \, \epsilon = \pm 1 \, .
\end{displaymath}
Thus, one obtains the following theorem.

\begin{theorem} {\rm \cite{FK}}
  \label{th:3}
  A complete list of superconformal algebras consists of loop
  algebras $\tilde{\fg}$, where $\fg$ is a simple
  finite-dimensional Lie superalgebra, and of Lie superalgebras
  $(N \in \ZZ_+)$:  $W_{(N)}$, $S'_{(N+2),\epsilon,a}$ ($N$ even and $a=0$ if
  $\epsilon =1$),
   $K_{(N)} (N \neq 4)$, $K'_{(4)}$, and $CK_{(6)}$.

\end{theorem}

Note that the first members of the above series are well-known
superalgebras:  $W_{(0)}\simeq K_{(0)}$ is the Virasoro algebra,
$K_{(1)}$ is the Neveu-Schwarz algebra, $K_{(2)}\simeq W_{(1)}$ is
the $N=2$ algebra, $K_{(3)}$ is the $N=3$ algebra,
$S'_{(2)}$=$S'_{(2),0,0}$ is the $N=4$ algebra, $K'_{(4)}$ is the
big $N=4$ algebra  (all centerless).  These algebras, along with
$W_{(2)}$ and $CK_{(6)}$ are the only superconformal algebras for
which all fields are primary with positive conformal weights
\cite{K6}.  It is interesting to note that all of them are
contained in $CK_{(6)}$, which consists of $32$~fields, the even
ones are the Virasoro fields and $15$~currents that form
$\tilde{so}_6$, and the odd ones are $6$ and $10$ fields of
conformal weight $3/2$ and $1/2$, respectively.  Here is the table
of (some) inclusions, where in square brackets the number of
fields is indicated:
\begin{eqnarray*}
  \begin{array}{cccc}
CK_{(6)} [32] & \supset & W_{(2)} [12] &
\supset W_{(1)} =K_{(2)}[4] \supset K_{(1)} [2] \supset \Vir [1]\\
\cup && \cup\\
K_{(3)} [8] \subset K'_{(4)}[16]\qquad && S'_{(2),\epsilon,a}[8]
  \end{array}\, .
\end{eqnarray*}
All of these Lie superalgebras have a unique non-trivial central
extension, except for $K'_4$ that has three \cite{KL} and
$CK_{(6)}$ that has none.  All other Lie superalgebras listed by
Theorem~\ref{th:2} have no non-trivial central extensions. (The
presence of a central term is necessary for the existence of
interesting representations and the construction of an interesting
conformal field theory.)

\section{Representations of affine superalgebras and ``almost''
  modular forms}
\label{sec:3}

\vskip-5mm \hspace{5mm}

\subsection{}
\label{sec:3.1}

Finite-dimensional irreducible representations of simple
finite-dimensional Lie superalgebras are much less understood than in
the Lie algebra case, the main reason being the occurence of isotropic
roots in the super case.  (A review may be found in the proceedings of
the last ICM, see \cite{Se}.)  The natural analogues of these
representations in the case of affine (super)algebras are the integrable
highest weight modules.

Let us first recall the basic definitions in the Lie algebra case,
i.e.,~for an affine Kac-Moody algebra $\hat{\fg}$ \cite{K3}.  Let
$\fg$ be a finite-dimensional simple Lie algebra and let $(.\,| \,
.)$ be an invariant symmetric bilinear form on $\fg$ normalized by
the condition that $(\alpha | \alpha)=2$ for a long root $\alpha$
($(a|b) = \tr ab$ in the case $\fg =s\ell_m$).  Recall that the
associated \emph{affine algebra} is
\begin{displaymath}
  \hat{\fg} = (\CC [x,x^{-1}] \otimes_{\CC} \fg)
      \oplus \CC K \oplus \CC D
\end{displaymath}
with the following commutation relations
 $ (a,b \in \fg \, ; \, m,n \in \ZZ$ and $a (m)$ stands for $x^m
 \otimes a)$:
 \begin{displaymath}
   \left[ a(m) , b(n)\right] = \left[ a,b\right]
        (m+n)+m\delta_{m,-n}(a|b)K \, , \,
   \left[ D,a(m)\right] = ma (m) , \left[ K,\hat{\fg}\right] =0 \, .
 \end{displaymath}
Note that the derived algebra $\hat{\fg}' $ is a central extension
(by $\CC K$) of the loop algebra $\tilde{\fg}$ that has made an
appearance in Section~2.  (It is also isomorphic to $\fg
(\hat{A})$, where $\hat{A}$ is the extended Cartan matrix of
$\fg$, cf.~Sec.~\ref{sec:1.7}.)  As we shall see, without central
extension one loses all interesting representations.  In any
irreducible $\hat{\fg}$-modle $V$ one has:  $K = kI_V$; the number
$k$ is called the \emph{level} of $V$.  The scaling element $D$ is
necessary for the convergence of characters.

Choose a Cartan subalgebra $\fh$ of $\fg$ and let $\fg = \fh
\oplus (\oplus_{\alpha \in \Delta} \fg_{\alpha})$ be the root
space decomposition, where $\fg_{\alpha}$ denotes the root space
attached to a root $\alpha \in \Delta \subset \fh^*$.  Let
$\hat{\fh} = \fh + \CC K + \CC D$ be the Cartan subalgebra of
$\hat{\fg}$, and, as before, let $\fg_{\alpha}(m) = x^m \otimes
\fg_{\alpha} $.  We extend the invariant bilinear form from $\fh$
to a symmetric bilinear form on $\hat{\fh}$ by letting $(\fh | \CC
K + \CC D)=0$, $(K|K)=(D|D)=0$, $(K|D)=1$, and identify
$\hat{\fh}$ with $\hat{\fh}^*$ via this form.

A $\hat{\fg}$-module $V$ is called \emph{integrable} if the
following two properties hold:

\begin{deflist}{M}
\setcounter{bean}{0}

  \item 
    $\hat{\fh}$ is diagonizable on $V$,

  \item 
    for each $\alpha \in \Delta$ and $m \in \ZZ$, $\fg_{\alpha}
    (m)$ is locally finite on $V$.

\end{deflist}

Choose a set of positive roots $\Delta^+ \subset \Delta$, and let
$\fn^+ = \oplus_{\alpha \in \Delta^+} \fg_{\alpha}$, $\hat{\fn}^+
= \fn^+ + \sum_{n \geq 1} x^n \otimes \fg$.  For each $\Lambda \in
\hat{\fh}^*$ one defines an \emph{irreducible
  highest weight module } $L (\Lambda)$ over $\hat{\fg}$ as the
(unique) irreducible $\hat{\fg}$-module for which there exists a
non-zero vector $v_{\Lambda}$ such that
\begin{displaymath}
  hv_{\Lambda} = \Lambda (h) v_{\Lambda} \hbox{  for all } h \in \hat{\fh} \, , \, \hat{\fn}_+ v_{\Lambda}=0 \, .
\end{displaymath}
Without loss of generality we shall let $\Lambda (D) =0$; then the
spectrum of $-D$ on $L (\Lambda)$ is $\ZZ_+$.

Integrable highest weight modules over affine Lie algebras (they
are automatically irreducible) attracted a lot of attention in the
past few decades both of mathematicians and of physicists (some
aspects of the theory are discussed in \cite{K3}, \cite{Wa2}.)
Here I will only mention some relevant to the talk facts.  First,
the level $k$ of such a module is a non-negative integer (and
$k=0$ iff $\dim L (\Lambda)=1$), and there is a finite number of
them for each $k$.  One of the most remarkable properties of these
modules is modular invariance, which I explain below.

\subsection{}
\label{sec:3.2}

Let us coordinatize $\hat{\fh}$ by letting
\begin{displaymath}
 ( \tau , z ,t) = 2\pi i (z-\tau D + tK) \, ,
\end{displaymath}
where $z \in \fh$, $\tau ,t \in \CC$, and define the
\emph{character} of the $\hat{\fg}$-module $L (\Lambda)$ by:
\begin{displaymath}
  ch_{\Lambda} (\tau ,z,t) =\tr_{L(\Lambda)} e^{2\pi i (z-\tau D + tK)} \, .
\end{displaymath}
If $L(\Lambda)$ is integrable, then $ch_{\Lambda} (\tau ,z,t)$ is
a holomorphic function for $(\tau ,z,t) \in \H \times \fh \times
\CC$, where $\H =\{  \tau \in \CC \, | \Im \tau >0 \}$.

Recall the following well-known action of $SL_2 (\ZZ)$ on $\H
\times \fh \times \CC$:
\begin{displaymath}
\left(
  \begin{array}{cc}
a & b\\ c & d
  \end{array}\right) \cdot
(\tau , z,t) = \left( \frac{a \tau +b}{c \tau +d} \, , \,
  \frac{z}{c\tau + d} \, , \, t -
  \frac{c(z|z)}{2(c\tau +d)} \right) \, .
\end{displaymath}
Then it turns out that there exists an explicit rational number
$m_{\Lambda}$, called \emph{modular anomaly} (see \cite{K3},
(12.7.5)) such that the normalized character $\chi_{\Lambda}
=e^{2\pi im_{\Lambda}\tau} ch_{\Lambda}$ of an integrable
$L(\Lambda)$ is invariant with respect to a congruence subgroup of
$SL_2 (\ZZ)$ (see \cite{K3}, Chapter~13).  If $ch_{\Lambda}$ has
this property, one says that $L(\Lambda)$ is \emph{modular
invariant}.

\subsection{}
\label{sec:3.3}

It turns out that $L(\Lambda)$ is modular invariant for a much wider
(than integrable) collection of $\Lambda$'s, called \emph{admissible},
defined by the condition \cite{KW1}, \cite{KW2}, \cite{K4}:
\begin{displaymath}
  2(\Lambda + \hat{\rho}|\alpha)/(\alpha | \alpha) \in \QQ
  \backslash \{ 0,-1,-2,\ldots \} \hbox{  for all  }
  \alpha \in \hat{\Delta}^+ \hbox{  such that  }
  (\alpha | \alpha) \neq 0 \, .
\end{displaymath}
  (The conjecture of
  Wakimoto and myself is that these are all modular invariant
  $L(\Lambda)$.)  Here $\hat{\Delta}^+$ is the set of positive roots of $\hat{\fg}$
corresponding to $\hat{\fn}^+ : \hat{\Delta}^+ = \Delta^+ \cup\{
\alpha + nK |\alpha \in \Delta , n \geq 1 \} \cup \{ nK|n \geq 1
\}$ , and $\hat{\rho} \in \hat{\fh}^*$ is a vector satisfying
$(\hat{\rho}|\alpha_i) = \tfrac{1}{2} (\alpha_i | \alpha_i)$ for
$i=0,1, \ldots ,r$,  where $\Pi =\{ \alpha_1 , \ldots ,\alpha_r
\}$ are simple roots of $\Delta^+$, $\hat{\Pi}=\{ \alpha_0
:=K-\theta \} \cup \Pi$ are simple roots of $\hat{\Delta}^+$,
$\theta$ is the highest root of $\Delta^+$.

I shall describe explicitly the most important class of them,
called \emph{principal admissible}.  Fix a positive integer $u$
and let $\hat{\Pi}_u = \{ uK-\theta \} \cup \Pi$.
Let $k=v/u$ be a rational number, such that $v \in \ZZ$ is
realtively prime to $u$ and $u(k+h\spcheck) \geq h\spcheck$.  Here
$h\spcheck$ is the dual Coxeter number (defined in a more general
Lie superalgebra context further on). Let $W$ be the Weyl group of
$\fg$ and let $P\spcheck = \{ \lambda \in \fh | (\lambda | \alpha)
\in \ZZ$ for all  $\alpha \in \Delta \}$. For each $\alpha \in
P\spcheck$ define a translation $t_{\alpha} \in \End \hat{\fh}$ by
the formula $  t_{\alpha} (\lambda) = \lambda + (\lambda |K)
\alpha -
  ((\lambda | \alpha)+ \tfrac{(\lambda |K)}{2} (\alpha | \alpha)) K$.
  Pick an
element $\hat{w} = t_{\beta} w$, where $w \in W$, such that
$\hat{w} \hat{\Pi}_u \subset \hat{\Delta}^+$ (these are all
subsets of $\hat{\Delta}^+$ isomorphic to $\hat{\Pi}$).  Let
$\Lambda^0$ be an integrable highest weight of level
$u(k+h\spcheck)-h\spcheck$. Then
\begin{displaymath}
\Lambda = \hat{w} (\Lambda^0 + \hat{\rho} - (u-1)
(k+h\spcheck)D)-\hat{\rho}
\end{displaymath}
is a \emph{principal admissible weight} of level $k$.  The
character of the corresponding $L(\Lambda)$ is given by the
following formula \cite{KW1}--\cite{KW2}:
%
\begin{gather}
  \left( \hat{R} ch_{L(\Lambda)}\right) (\tau ,z,t) =
    \left( \hat{R} ch_{L(\Lambda^0)}\right)
    \left( u\tau , w^{-1}(z+\tau\beta) \, , \,
    u^{-1} \left( t+ (z|\beta)+\tfrac{1}{2} \tau (\beta |\beta)\right)\right)
      \, ,
\tag{Ch}
\end{gather}
%
where $\hat{R} =e^{\hat{\rho}} \prod_{\alpha \in \hat{\Delta}^+}
(1-e^{-\alpha})^{\mult \alpha}$ is the Weyl denominator function
for $\hat{\fg}$, and $\mult \alpha =1$ except for $\mult nK =r$
for all $n$  (note that this formula is a tautology if $\Lambda =
\Lambda^0$ is integrable; this happens iff $u=1$).  Recall that
$ch_{L(\Lambda^0)}$ is given by the Weyl-Kac character formula
\cite{K3}.

\subsection{}
\label{sec:3.4}

Let now $\fg = \fg (A)$ be a basic simple finite-dimensional Lie
superalgebra (see Sec.~1.7).  Then $\fg$ carries a unique, up to a
constant factor, non-degenerate invariant bilinear form $B$
(``invariant'' means that $B ([a,b],c) = B (a,[b,c])$).  Let $\fh =
\sum^r_{i=1} \CC h_i$ be the Cartan subalgebra of $\fg$, $\fn^+$ the
subalgebra of $\fg$ generated by all $e_i$, $\Delta^+ \subset \fh^*$ the
set of positive roots (i.e.,~roots of $\fh$ in $\fn^+$), $\Delta =
\Delta^+ \cup - \Delta^+$ the set of all roots, $\Delta_{\bar{0}}$ and
$\Delta_{\bar{1}}$ the sets of even and odd roots, $\{ \alpha_1 , \ldots
, \alpha_r \} \subset \Delta^+$ the set of simple roots $(\alpha_i (h_j)
= a_{ji})$, $\theta \in \Delta^+$ the highest root.  Define $\rho \in
\fh^*$ by $B (\rho ,\alpha_i) = \tfrac{1}{2} B (\alpha_i , \alpha_i)$,
$i=1,\ldots , r$.  Then $  h\spcheck_B = B (\rho , \rho) + \tfrac{1}{2}
B (\theta , \theta)$ is the eigenvalue of the Casimir operator in the
adjoint representation.

If $\fh\spcheck_B \neq 0$, we let $ \Delta^\#_0 = \{ \alpha \in
\Delta_{\bar{0}}\, |\,  h\spcheck_B B (\alpha , \alpha)> 0 \}$. If
$h\spcheck_B =0$, which happens for $\fg =s\ell (m|m)/\CC E$, $spo
(2m|2m+2)$ and $D(2,1;a)$, we take for $\Delta^{\#}_0$ the sets of
roots of the subalgebra $s\ell_m$, $so_{2m+2}$ and $s\ell_2 \oplus
s\ell_2$, respectively.  Let $W^{\#}$ denote the subgroup of the
Weyl group of $\fg_{\bar{0}}$ generated by reflections with
repsect to all $\alpha \in \Delta^{\#}_0$.  Denote by $(. \, | \,
.)$ the invariant bilinear form on $\fg$ normalized by the
condition $  (\alpha | \alpha) =2$ for the longest root $\alpha
\in \Delta^{\#}_0$. The corresponding to this form number
$h\spcheck = h\spcheck_{(. \, | \, .)}$ is called \emph{the dual
Coxeter
  number}.  (For example, this number equals $|m-n|$ for $s\ell
(m|n), \tfrac{1}{2}(m-n)+1$ for $spo (m|n)$ with $m\geq n-2$, $30$
for $E_8$, $3$ for $F(4)$, $2$ for $G(3)$.)

\subsection{}
\label{sec:3.5}

Define the affine superalgebra $\hat{\fg}$ associated to the Lie
superalgebra $\fg$ in exactly the same way as in the Lie algebra
case. The highest weight $\hat{\fg}$-modules $L(\Lambda)$ are
defined in the same way too.  The \emph{integrability} of a
$\hat{\fg}$-module $V$ is defined by (M1), (M2) with $\Delta$
replaced by $\Delta^{\#}_0$, and

\begin{deflist}{M}
\setcounter{bean}{2}

  \item 
    $V$ is locally $\fg$-finite.  ((M1) and (M2) imply (M3), but
    there are no integrable $L(\Lambda)$ in the super case if one
    doesn't replace $\Delta$ by $\Delta^{\#}_0$.)

\end{deflist}

Integrable $\hat{\fg}$-modules $L(\Lambda)$ were classified in
\cite{KW4}.  However, very little is known about their characters.
The following example shows that modular invariance fails already
in the simplest case $\fg = s\ell (2|1), \Lambda =D$.  In this
case
\begin{gather}
  e^{-D} ch_{L(D)}= \prod^{\infty}_{n=1}
  ((1-q^n)^{-1}) (1+z_1q^\fn) (1+z^{-1}_1 q^{\fn-1})
     A (z^{-1}_1, z_2 q^{1/2},q) \, ,
\tag{A}
\end{gather}
where $z_i=e^{\epsilon_i +\epsilon_3}$ ($i=1,2$; $\epsilon_i$ is
the standard basis of the space of $3 \times 3$ diagonal
matrices), $q=e^{2\pi i \tau}$, and
\begin{displaymath}
  A(x,z,q) = \sum_{n \in \ZZ} \frac{q^{n^2/2} z^{n}
    }{1+xq^{n}} \, .
\end{displaymath}
The function $A(x,z,q)$ converges to a meromorphic function in the
domain $x,y,z \in \CC$, $|q|<1$, and is called Appell's function.
Since the first factor in (A) has the modular invariance property
and $A (x,z,q)$ doesn't have it, we see that $L(D)$ is not modular
invariant.  We call the Appell function an \emph{almost modular
form} since it is a section of a rank~$2$ vector bundle on an
elliptic curve for each $\tau \in \H$ \cite{P} (whereas modular
forms are sections of rank~$1$ vector bundles on it).

We call a weight $\Lambda \in \hat{\fh}^*$ of the Lie superalgebra
$\hat{\fg}$ \emph{admissible} (resp. \emph{ principal admissible})
 if $\Lambda $ is admissible (resp. principle admissible) for
the affine Lie algebra associated to a semisimple Lie algebra with
root system $\Delta^{\#}_0$,  and we conjecture \cite{KW4} that
formula (Ch) still holds, where $\hat{R}$ in the super case is
defined by $\hat{R} = e^{\hat{\rho}}\prod_{\alpha \in
\hat{\Delta}^+} (1-(-1)^{p(\alpha)} e^{-\alpha})^{(-1)^{p(\alpha)}
\mult
  \alpha}$. Unfortunately, $ch_{L(\Lambda^0)}$ is not known in
general, but in the \emph{boundary level} case, i.e.,~when
$u(k+h\spcheck) = h\spcheck$, the level of $\Lambda^0$ is zero,
hence $ch_{L(\Lambda^0)}=1$, and formula~(Ch) gives an explicit
expression for $ch_{L(\Lambda)}$.  In particular, the character is
modular invariant in this case.

Let me mention in conclusion of this section that the Weyl-Kac
character formula in the case of $1$-dimensional module over an
affine Lie algebra $\hat{\fg}$ turns into celebrated Macdonald's
identities that express $\hat{R}$ as an infinite series, the
special case for $\fg = s\ell_2$ being the Jacobi triple product
identity.  A sum formula for $\hat{R}$ in the super case is also
known \cite{KW3} (see also the talk \cite{Wa1} at the last ICM).
In the simplest case of $\fg = s\ell (2|1)$ one gets the identity:
\begin{displaymath}
  \prod^{\infty}_{n=1} \frac{(1-q^n)^2
    (1-uvq^{n-1})(1-u^{-1}v^{-1}q^n)}
  {(1-uq^{n-1})(1-u^{-1}q^n) (1-vq^{n-1}) (1-v^{-1}q^n)}
  = \left( \sum^{\infty}_{m,n=0} -  \sum^{-\infty}_{m,n=-1} \right)
    u^mv^n q^{mn}\, ,
\end{displaymath}
which goes back to Ramanujan and even further back to Kronecker.

\section{Quantum reduction for affine Lie superalgebras}
\label{sec:4}

\vskip-5mm \hspace{5mm}

\subsection{}
\label{sec:4.1}

This part of my talk is based on a joint work with S.-S. Roan and
M. Wakimoto \cite{KRW}, \cite{KW5}.  I will explain a general
quantum reduction scheme, which is a further development of a
number of works.  They include \cite{DS}, \cite{KS} and \cite{Kh}
on classical Drinfeld-Sokolov reduction, and \cite{FF1},
\cite{FF2}, \cite{FKW}, \cite{B}, \cite{BT} on its quantization.
As in \cite{FF2}, the basic idea is to translate the geometric
Drinfeld-Sokolov reduction to a homological language as in
\cite{KS}, and then to quantize this homology complex. Remarkably,
this procedure gives, starting from affine superalgebras, a number
of very interesting super extensions of the Virasoro algebra and
their most interesting representations. I will use the very
convenient language of vertex algebras (introduced in \cite{B}),
all related notations can be found in \cite{K5}.

\subsection{}
\label{sec:4.2}

Let $\fg$ be a basic simple finite-dimensional Lie superalgebra
with a non-degenerate invariant bilinear form $(. \, | \, .)$. Fix
a number $k$ and a nilpotent even element $f$ of $\fg$. Include
$f$ in an $s\ell_2$-triple $\{ e,h,f\}$ so that $[h,e]=2e$,
$[h,f]=-2f$, $[e,f]=h$.  We have the eigenspace decomposition of
$\fg$ with respect to $\ad h$:  $\fg = \oplus_{j \in \ZZ} \fg_j$,
and we let $\fg_+ = \oplus_{j>0} \fg_j$.  The element $f$ defines
a non-degenerate skew-supersymmetric bilinear form $\langle . \, ,
\, . \rangle$ on $\fg_1$ by the formula $\langle a ,b \rangle =
(f| [a,b])$.  Let $A_{ne}$ denote the superspace $\fg_1$ with the
form $\langle . \, , \, . \rangle$. Denote by $A_{ch}$ the
superspace $\pi \fg_+ + \pi \fg^*_+$, where $\pi$ stands for the
reversal of parity, with the skew-supersymmetric bilinear form
defined by $( a , b^* ) = b^* (a)$ for $a \in \pi\fg_+, b \in \pi
\fg^*_+$, $( \pi \fg_+ , \pi\fg_+ ) =0 =( \pi \fg^*_+ , \pi\fg^*_+
) $.

Let $V^k (\hat{\fg})$ be the universal affine vertex algebra, and
let $F^1 (A_{ne})$, $F^1(A_{ch})$ be the free fermionic vertex
algebras (\cite{K5}, \S~4.7).  Consider the vertex algebra
\begin{displaymath}
  C(\fg ,f,k) = V^k (\hat{\fg}) \otimes F^1 (A_{ne})
  \otimes F^1 (A_{ch}) \, ,
\end{displaymath}
and define its charge decomposition $C(\fg ,f,k)=\oplus_{m \in
  \ZZ} C_m$ by letting charge $V^k (\hat{\fg}) =$ charge $F^1
(A_{ne})=0$, charge $\pi \fg_+ =1 =-$ charge $\pi \fg^*_+$.

Next, we define a differential $d$ on $C(\fg ,f,k)$ which makes it
a homology complex.  For this choose a basis $\{ u_i \}_{i \in
  S'}$ of $\fg_1$ and extend it to a basis $\{ u_i \}_{i \in S}$
of $\fg_+$ compatible with its $\ZZ$-gradation.  Denote by $\{
\varphi_i \}_{i \in S}$ and $\{ \varphi^*_i \}_{i \in S}$ the
corresponding dual bases of $\pi \fg_+$ and $\pi \fg^*_+$, and by
$\{ \Phi_i \}_{i \in S'}$ the corresponding basis of $A_{ne}$.
Consider the following odd field of the vertex algebra $C(\fg
,f,k)$:
\begin{eqnarray*}
  d(z) &=& \sum_{i \in S} (-1)^{p(u_i)} u_i (z) \otimes
  \varphi^*_i (z) \otimes 1\\
  & & -\tfrac{1}{2} \sum_{i,j,k \in S} (-1)^{p(u_i)p(u_k)}
     c^k_{ij} \otimes \varphi_k (z) \varphi^*_i (z)
     \varphi^*_j (z) \otimes 1\\
     && + \sum_{i \in S} (f|u_i)  \otimes \varphi^*_i (z)
        \otimes 1 +\sum_{i \in S'} 1 \otimes \varphi^*_i
        (z) \otimes \Phi_i (z) \, ,
\end{eqnarray*}
where $[u_i,u_j]=\sum_{k} c^k_{ij} u_k$ in $\fg_+$, and let
$d=\Res_{z=0} d(z)$.  (Note that the first two summands of $d$
form the usual differential of a Lie (super)algebra complex.) Then
one checks that $[d(z), d(w)]=0$, hence $d^2=0$.  It is also clear
that $dC_m\subset C_{m-1}$.  We define the vertex algebra $W (\fg
,f, k)$ as the $0$\st{th} homology of this complex, and call it
the \emph{quantum reduction} of the triple $(\fg ,f,k)$ (actually,
it depends only on $\fg ,k$ and the conjugacy class of $f$ in
$\fg_{\bar{0}}$).

One of the fields of the vertex algebra $W (\fg ,f,k)$ is the
following Virasoro field
\begin{eqnarray*}
  L(z) &=& \frac{1}{2(k+h\spcheck)} \sum_i :
    a_i (z) b_i (z): + \frac{1}{2}\partial_z h(z) + \sum_{i \in S}
    ((1-m_i):\partial_z \varphi^*_i (z) \varphi_i (z) : \\
    &&  -m_i : \varphi^*_i (z) \partial_z \varphi_i (z) :)
    + \tfrac{1}{2} \sum_{i \in S'} g^{ij}:\partial_z \Phi_i
    (z) \Phi_j (z): \, ,
\end{eqnarray*}
where $\{ b_i \}$ and $\{a_i \}$ are dual bases of $\fg$:
$(b_i|a_j)=\delta_{ij}$ , $[h,u_i]=2m_iu_i$, $(g^{ij})$ is the
matrix inverse to $(\langle u_i,u_j \rangle)_{i,j \in S'}$. The
central charge of $L(z) $ is equal to:
\begin{displaymath}
  c(k) = \frac{k \sdim \fg}{k+h\spcheck} - 3 (h|h)k
  -\sum_{i \in S}(-1)^{p(u_i)} (12 m^2_i - 12m_i+2)
  - \tfrac{1}{2} \sdim \fg_1 \, .
\end{displaymath}
Here $\sdim V = \dim V_{\bar{0}} -\dim V_{\bar{1}}$ is the
superdimension of the superspace $V$.  Thus, all $W(\fg ,f,k)$ are
super-extensions of the Virasoro algebra.  Furthermore, for each
$\ad h$ eigenvector with eigenvalue $-2j$ in the centralizer of
$f$
 in $\fg$, $W(\fg ,f,k)$ contains a field of conformal weight
 $1+j$ (so that $L(z)$ corresponds to $f$), and these fields
 generate the vertex algebra $W(\fg ,f,k)$.

\vspace{1ex}

\begin{examples*}
   \arabicparenlist
   \begin{examenum}
\arabicparenlist
%
\item
  $\fg$ is a simple Lie algebra.
\alphaparenlistii
\begin{examenum}
\item 
  $f$ is a principal nilpotent element.  Then $W (\fg ,f,k)$ is
called the quantum Drinfeld-Sokolov reduction.  These algebras and
their representations were extensively studied in \cite{FF2},
\cite{FB}, \cite{FKW} and many other papers.  The simplest case of
$\fg = s\ell_2$ produces the Virasoro vertex algebra.  The case
$\fg = s\ell_3$ gives the $W_3$ algebra \cite{Z}.

\item 
$f$ is a lowest root vector of $\fg$.  These vertex algebras were
discussed from a different point of view in \cite{FL} under the
name quasi-superconformal algebras.  The special case of $\fg =
s\ell_3$ was studied from a quantum reduction viewpoint in
\cite{Be}.

\end{examenum}

\item
  $\fg$ is a simple basic Lie superalgebra and $f$ is an
  even lowest root vector.

 \alphaparenlistii
\begin{examenum}
\item 
One has the following correspondence:

  \begin{center}
\begin{tabular}{c|c|c|c|c|c}
$\fg $ & $spo (2|1)$ & $s\ell (2|1)$ & $s\ell (2|2)$
       & $ spo (2|3)$ & $D(2,1;a)$\\[1ex]
\hline
$W(\fg ,f,k)$ & Neveu-Schwarz & $N=2$ & $N=4$ & $ N=3$ & big $N=4$
\end{tabular}
  \end{center}
%
(In the last two columns one gets an isomorphism after adding one
fermion, resp. four fermions and one boson.).

\item 
Almost every lowest root vector of a simple component of
$\fg_{\bar{0}}$ can be made equal $f$.  This gives all
superconformal algebras of \cite{FL} (by definition, they are
generated by the Virasoro field, the even fields of weight~$1$ and
$N$ odd fields of weight $3/2$), and many new examples.

\end{examenum}
\end{examenum}
\end{examples*}

\subsection{}
\label{sec:4.3}

Let $M$ be a highest weight module over $\hat{\fg}$.  It extends
to a vertex algebra module over $V^k(\hat{\fg})$, and we consider
the $C (\fg ,f,k)$-module $C (M) = M \otimes F^1 (A_{ne}) \otimes
F^1 (A_{ch})$. The element $d$ acts on $C(M)$ and again $d^2=0$,
hence we can consider homology $H(M) = \oplus_j H_j (M)$ which is
a module over $W (\fg ,f,k)$.  The $W(\fg ,f,k)$-module $H(M)$ is
called the \emph{quantum reduction} of the $\hat{\fg}$-module $M$.
Using the Euler-Poincar\'e principle one easily computes the
character of $H(M)$ in terms of $ch_M$.  The basic conjecture of
\cite{FKW}, \cite{KRW} is that $H(M)$ is irreducible (in
particular, at most one $H_j (M)$ is non-zero) if $M$ is
admissible.

   \begin{examples*}
   \arabicparenlist
   \begin{examenum}

\item
  $\fg = s\ell_2$.  Let $k$ be an admissible level, i.e.,~$k$ is
  a rational number with positive denominator $u$ such that
  $u(k+2) \geq 2$ (recall that $h\spcheck =2$).  The set of
  principal admissible weights of level $k$ is as follows ($\alpha$ is a
  simple root of $s\ell_2$) \cite{KW1}, \cite{K4}:
  \begin{displaymath}
    \{ \Lambda_{k,j,n} = kD +\tfrac{1}{2} (n-j(k+2)) \alpha \, | \,
    0 \leq j \leq u-1 \, , \,
    0 \leq n \leq u(k+2)-2 \} \, .
  \end{displaymath}
Then the quantum reduction of the $\hat{s\ell}_2$-module
$L(\Lambda_{k,j,n})$ is the ``minimal series'' module
corresponding to parameters $p=u(k+2)$, $p'=u$ (cf.~\cite{BPZ},
\cite{K4}):
\begin{displaymath}
  c^{(p,p')} =1-6\frac{(p-p')^2}{pp'} \, , \,
   h^{(p,p')}_{j+1,n+1}  = \frac{(p(j+1)-p'(n+1))^2 -(p-p')^2}
   {4pp'} \, .
\end{displaymath}
The character formula~(Ch) for $L(\Lambda_{k,j,n})$ gives
immediately all the characters of minimal series.

\item 
  $\fg =spo (2|1)$.  We get all minimal series modules
  over the Neveu-Schwarz algebra and their characters by quantum
  reduction of all (not only principal) admissible $\hat{\fg}$-modules.

\item
  $\fg = s\ell (2|1)$.  Then the boundary admissible levels are $
 k=m^{-1}-1$, where $m \in \ZZ$, $m\geq 2$.  One has the following
 $\hat{w} \hat{\Pi}_m$'s (see Sec.~\ref{sec:3.3}), where  $\alpha_1$
 and $\alpha_2$ are odd:
$
   \{ \alpha_0 +b_0K \, , \,
   \alpha_1 + b_1K \, , \,
   \alpha_2 + b_2K \} $.
Here $b_i$ are non-negative integers, $b_0 \geq 1$ and $\sum b_i
=m-1$. The quantum reduction of the corresponding admissible
$\hat{\fg}$-modules  gives all the minimal series representations
of the $N=2$ superconformal algebra (cf.~\cite{FST} and references
there). Again, formula~(Ch) gives immediately their characters.

\item 
$\fg = s\ell (2|2)$ (resp. $spo (2|3)$).  In a similar fashion we
recover the characters of $N=4$ \cite{ET} (resp. $N=3$ \cite{M})
superconformal algebras.

 \end{examenum}

   \end{examples*}

\section{\boldmath Representations of $E(3|6)$ and the standard model}
\label{sec:5}

\vskip-5mm \hspace{5mm}

\subsection{}
\label{sec:5.1}

By a representation of a linearly compact Lie superalgebra $L$ we shall
mean a continuous representation in a vector space $V$ with discrete
topology (then the contragredient representation is a continuous
representation in a linearly compact space $V^*$). Fix an open
subalgebra $L_0$ of $L$.  We shall assume that $V$ is locally
$L_0$-finite, meaning that any vector of $V$ is contained in a
finite-dimensional $L_0$-invariant subspace (this property actually
often implies that $V$ is continuous).  These kinds of representations
were studied in the Lie algebra case by Rudakov \cite{R}.  It is easy to
show that such an irreducible $L$-module $V$ is a quotient of an induced
module $\Ind^L_{L_0}U=U(L) \otimes_{U(L_0)}U$, where $U$ is a
finite-dimensional irreducible $L_0$-module, by a (unique in good cases)
maximal submodule.  The induced module $\Ind^L_{L_0}U$ is called
\emph{degenerate} if it is not irreducible.  An irreducible quotient of
a degenerate induced module is called a \emph{degenerate} irreducible
module.

One of the most important problems of representation theory is to
determine all degenerate representations. I will state here the
result for $L=E(3|6)$ with $L_0=\Pi_{j \geq 0} \fa_j$
(see~Sec.~\ref{sec:1.6}).  The finite-dimensional irreducible
$L_0$-modules are actually $\fa_0=s\ell_3 \oplus s\ell_2 \oplus
g\ell_1$-modules (with $\Pi_{j>0}\fa_j$ acting trivially).  We
shall normalize the generator $Y$ of $g\ell_1$ by the condition
that its eigenvalue on $\fa_{-1}$ is $-1/3$.  The
finite-dimensional irreducible $\fa_0$-modules are labeled by
triples $(p,q;r;Y)$, where $p,q$ (resp.~$r$) are labels of the
highest weight of an irreducible representation of $s\ell_3$
(resp.~$s\ell_2$), so that $p,0$ and $0,q$ label $S^p\CC^3$ and
$S^q\CC^{3*}$ (resp. $r$ labels $S^r\CC^2$), and $Y$ is the
eigenvalue of the central element $Y$. Since irreducible
$E(3|6)$-modules are unique quotients of induced modules, they can
be labeled by the above triples as well.

\begin{theorem} {\rm \cite{KR1}--\cite{KR3}}
  \label{th:4}
  The complete list of irreducible degenerate $E(3|6)$-modules
consists of four series:  $ (p,0;r;-r+\tfrac{2}{3}p) ,
 (p,0;r;r+\tfrac{2}{3}p+2)\,  , \,    (0,q;r;-r-\tfrac{2}{3}q-2) ,
 (0,q;r;r-\tfrac{2}{3} q)$.

\end{theorem}

\subsection{}
\label{sec:5.2}

Remarkably, all four degenerate series occur as cokernels of the
differential of a differential complex $(M,\triangledown)$ constructed
below (see \cite{KR2} for details).  I shall view $E(3|6)$ as a
subalgebra of $E(5|10) = \Pi_{j \geq -2} \fg_j$ as in
Sect.~\ref{sec:1.6}, expressed in terms of vector fields and
  differential forms in the indeterminates $x_1$, $x_2$, $x_3$,
$  z_+ = x_4$, $z_- = x_5$.  Recall that $\fg_0$ is the algenra of
divergenceless vector fields with linear coefficients.  Let
$Y=\tfrac{2}{3} \sum_i x_i \partial_i -\sum_{\epsilon}
z_{\epsilon} \partial_{\epsilon} \in \fg_0$.  Here and further
$i=1$, $2$, $3$, $\epsilon =+,-$, and $\partial_i =\partial
/\partial x_i$, $\partial_{\epsilon} = \partial /\partial
z_{\epsilon}$.  Then $\fa_0$ is the centralizer of $Y$ in $\fg_0$
and $\fa_{-1}$ is the span of all elements $d^{\epsilon}_i =dx_i
\wedge dz_{\epsilon}$.

Consider the following four $\fa_0$-modules (extended to
$L_0$-modules by trivial action of $\fa_j$ with $j >0$):
\begin{displaymath}
  V_I = \CC [x_i , z_{\epsilon}] \, , \,
  V_{II} = \CC [x_i , \partial_{\epsilon}]_{[2]} \, , \,
  V_{III} = \CC [\partial_i ,z_{\epsilon}]_{[-2]} \, ,\,
  V_{IV}   = \CC [\partial_i , \partial_{\epsilon}] \, ,
\end{displaymath}
where the subscript $[a]$ means that $Y$ is shifted by the scalar
$a$.  For each $R=I-IV$ introduce a bigradation $V_R = \oplus_{m,n
\in \ZZ} V^{(m,n)}_R$ by letting $\deg x_i = (1,0)$, $\deg
z_{\epsilon} = (0,1)$, and let $M_R = \Ind^L_{L_0} V_R
=\oplus_{m,n \in \ZZ} M^{(m,n)}_R$.  Then the non-zero
$M^{(m,n)}_R$ are all the degenerate $E(3|6)$-modules of the
$R$\st{th} series.  We let
\begin{displaymath}
  M= \left( \oplus_{(m,n) \neq (0,0)} M^{(m,n)}_I \right)
\oplus M_{II} \oplus M_{III} \oplus \left( \oplus_{(m,n)\neq
(0,0)} M^{(m,n)}_{IV} \right) \, .
\end{displaymath}
The differentials $\triangledown_k$ introduced further are
elements of $U(L) \otimes \End V$ that act on
$U(L)\otimes_{U(L_0)}V$ by the formula:
\begin{displaymath}
  ( \sum_j u_j \otimes A_j ) (u \otimes v)
    = \sum_j uu_j \otimes A_j v \, .
\end{displaymath}


\noindent \textbf{Example.}\quad The dual to the ordinary formal
de Rham complex is \linebreak $\CC [\partial /\partial x_1 ,
\ldots ,
\partial /\partial x_m] \otimes \Lambda (\partial / \partial \xi_1
, \ldots , \partial / \partial \xi_m)$ with the differential $d^*
=\sum_j
\partial / \partial x_j \otimes \xi_j$ and the $\ZZ$-gradation
defined by $\deg \partial /
\partial x_i =0$, $\deg \partial / \partial \xi_i =1$.  Rudakov's
theorem \cite{R} says that all irreducible degenerate
$W_m$-modules occur as cokernels of $d^*$. \vspace{2ex}

Turning now to $\triangledown_k$, we let $\Delta^{\pm} = \sum_i
d^{\pm}_i \otimes \partial_i$, $\delta_i = d^+_i \otimes
\partial_+ + d^-_i \otimes \partial_-$.  Then $\triangledown_1 =
\Delta^+ (1 \otimes \partial_+) + \Delta^- (1 \otimes
\partial_-)$ is a well-defined operator on all $M_R$ such that
$\triangledown^2_1=0$.  Furthermore there are differentials
$\triangledown_2 = \Delta^+ \Delta^-$, $\triangledown_3 =
\delta_1\delta_2\delta_3$, $\triangledown_4$, $\triangledown'_4$
and $\triangledown_6$ (the explicit expressions of the last three
can be found in \cite{KR2}) that sew together these four
complexes.  These differentials are illustraed by Table~M.  The
white nodes and black marks represent the induced modules of the
$R$\st{th} series.  The plain arrows represent $\triangledown_1$,
the dotted arrows represent $\triangledown_2$, the interrupted
arrows represent $\triangledown_3$ and the bold arrows represent
$\triangledown'_4$, $\triangledown''_4$ and $\triangledown_6$. The
white nodes denote the places with zero homology.  The black marks
denote the places with non-zero homology, also computed in
\cite{KR2}.  For example, at the star mark the homology is $\CC$.

Similar results for $E(3|8)$ and (to a lesser extent) for
$E(5|10)$ are given in \cite{KR4}.

\begin{table}[htbp]
  \begin{center}
    \leavevmode
  \setlength{\unitlength}{0.25in}
  \vspace{0.5ex}
\begin{picture}(21,18.5)

\put(18.7,17.5){I} \put(18.7,-1){II}

\put(1,17.5){III} \put(1,-1){IV}

\put(8.5,17.5){\line(0,-1){7} }
\put(8.4,18){r}

\put(11.5,17.5){\line(0,-1){8.5} }
\put(11.4,18){r}

\put(10,-.5){\line(0,1){9.5} }
\put(9.9,-1){r}

\put(13,-.5){\line(0,1){8} }
\put(12.9,-1){r}

\put(13,7.5){\line(1,0){5} }
\put(18.5,7.4){p}

\put(1.8,9){\line(1,0){8.2} }
\put(11.5,9){\line(1,0){6.5} }
\put(1,8.9){q} \put(18.5,8.9){p}

\put(1.8,10.5){\line(1,0){6.8} }
\put(1,10.4){q}

\thicklines
\multiput(2.5,0)(0,1.5){12}{\circle{.25} }
\multiput(4,0)(0,1.5){12}{\circle{.25} }
\multiput(5.5,0)(0,1.5){12}{\circle{.25} }
\multiput(7,0)(0,1.5){12}{\circle{.25} }

\multiput(8.5,0)(0,1.5){4}{\circle{.25} }
\multiput(8.5,9)(0,1.5){6}{\circle{.25} }
\multiput(8.3,5.9)(0,1.5){1}{$\bigstar$ }
\multiput(8.4,7.4)(0,1.5){1}{$\spadesuit$ }

\multiput(10,0)(0,1.5){6}{\circle{.25} }

\multiput(11.5,10.5)(0,1.5){5}{\circle{.25} }

\multiput(13,0)(0,1.5){7}{\circle{.25} }
\multiput(13,12)(0,1.5){4}{\circle{.25} }
\multiput(12.8,10.4)(0,1.5){1}{$\blacklozenge $}

\multiput(14.5,0)(0,1.5){12}{\circle{.25} }
\multiput(16,0)(0,1.5){12}{\circle{.25} }
\multiput(17.5,0)(0,1.5){12}{\circle{.25} }
\thinlines

\thinlines
\drawline(1.95,-.55)(2.4,-.1)
\drawline(1.95, .95)(2.4,1.4)
\drawline(1.95,2.45)(2.4,2.9)
\drawline(1.95,3.95)(2.4,4.4)
\drawline(1.95,5.45)(2.4,5.9)
\drawline(1.95,6.95)(2.4,7.4)
\drawline(1.95,8.45)(2.4,8.9)

\drawline(1.95,11.45)(2.4,11.9)
\drawline(1.95,12.95)(2.4,13.4)
\drawline(1.95,14.45)(2.4,14.9)
\drawline(1.95,15.95)(2.4,16.4)

\drawline(3.45,-.55)(3.9,-.1)
\drawline(4.95,-.55)(5.4,-.1)
\drawline(6.45,-.55)(6.9,-.1)
\drawline(7.95,-.55)(8.4,-.1)
\drawline(9.45,-.55)(9.9,-.1)

\drawline(13.95,-.55)(14.4,-.1)
\drawline(15.45,-.55)(15.9,-.1)
\drawline(16.95,-.55)(17.4,-.1)

\multiput(18.2,.7)(0,1.5){12}{\vector(-1,-1){.55}}
\multiput(3.2,17.2)(1.5,0){5}{\vector(-1,-1){.55}}
\multiput(12.2,17.2)(1.5,0){5}{\vector(-1,-1){.55}}


\multiput(3.85,16.35)(1.5,0){4}{\vector(-1,-1){1.2}}
\multiput(12.85,16.35)(1.5,0){4}{\vector(-1,-1){1.2}}

\multiput(3.85,14.85)(1.5,0){4}{\vector(-1,-1){1.2}}
\multiput(12.85,14.85)(1.5,0){4}{\vector(-1,-1){1.2}}

\multiput(3.85,13.35)(1.5,0){4}{\vector(-1,-1){1.2}}
\multiput(12.85,13.35)(1.5,0){4}{\vector(-1,-1){1.2}}

\multiput(3.85,11.85)(1.5,0){4}{\vector(-1,-1){1.2}}
\multiput(12.85,11.85)(1.5,0){4}{\vector(-1,-1){1.2}}

\dashline[-10]{.1}(3.85,10.35)(2.35,8.85)
       \put(7.33,9.37){\vector(-1,-1){.25}}
\dashline[-10]{.1}(5.35,10.35)(3.85,8.85)
       \put(5.8,9.37){\vector(-1,-1){.25}}
\dashline[-10]{.1}(6.85,10.35)(5.35,8.85)
       \put(4.3,9.37){\vector(-1,-1){.25}}
\dashline[-10]{.1}(8.35,10.35)(6.85,8.85)
       \put(2.9,9.35){\vector(-1,-1){.25}}

\multiput(14.35,10.35)(1.5,0){3}{\vector(-1,-1){1.2}}

\multiput(3.85,8.85)(1.5,0){4}{\vector(-1,-1){1.2}}
\dashline[-10]{.1}(14.35,8.85)(12.85,7.35)
\dashline[-10]{.1}(15.85,8.85)(14.35,7.35)
\dashline[-10]{.1}(17.35,8.85)(15.85,7.35)
\dashline[-10]{.1}(2.25,10.25)(2,10)

      \put(13.4,7.9){\vector(-1,-1){.25}}
      \put(14.9,7.9){\vector(-1,-1){.25}}
      \put(16.4,7.9){\vector(-1,-1){.25}}


\dashline[-10]{.2}(10.3,16.8)(8.5,15)
       \put(8.8,15.35){\vector(-1,-1){.25}}

\dashline[-10]{.2}(11.5,16.5)(8.5,13.5)
       \put(8.8,13.85){\vector(-1,-1){.25}}

\dashline[-10]{.2}(11.5,15)(8.5,12)
       \put(8.8,12.35){\vector(-1,-1){.25}}

\dashline[-10]{.2}(11.5,13.5)(8.5,10.5)
       \put(8.8,10.85){\vector(-1,-1){.25}}


\thicklines \put(11.5,12){\vector(-1,-1){2.81}}
\put(13,9){\vector(-1,-1){2.81}}

\put(11.49,10.49){\vector(-1,-2){1.44}}
\put(11.51,10.51){\line(-1,-2){1.41}}
\put(11.47,10.47){\line(-1,-2){1.41}}

\dashline[-5]{.3}(12.8,10.5)(9.1,7.9)
      \put(9.1,7.9){\vector(-3,-2){.3}}
\thinlines

\dashline[-10]{.2}(13,7.5)(10,4.5)
      \put(10.4,4.9){\vector(-1,-1){.25}}

\dashline[-10]{.2}(13,6)(10,3)
      \put(10.4,3.4){\vector(-1,-1){.25}}

\dashline[-10]{.2}(13,4.5)(10,1.5)
      \put(10.4,1.9){\vector(-1,-1){.25}}

\dashline[-10]{.2}(13,3)(10,0)
     \put(10.4,.4){\vector(-1,-1){.25}}



\dashline[-10]{.2}(13,1.5)(11,-.5)
\dashline[-10]{.2}(13,0)(12.5,-.5)


\multiput(3.85,7.35)(1.5,0){5}{\vector(-1,-1){1.2}}
\multiput(14.35,7.35)(1.5,0){3}{\vector(-1,-1){1.2}}
\multiput(3.85,5.85)(1.5,0){5}{\vector(-1,-1){1.2}}
\multiput(14.35,5.85)(1.5,0){3}{\vector(-1,-1){1.2}}
\multiput(3.85,4.35)(1.5,0){5}{\vector(-1,-1){1.2}}
\multiput(14.35,4.35)(1.5,0){3}{\vector(-1,-1){1.2}}
\multiput(3.85,2.85)(1.5,0){5}{\vector(-1,-1){1.2}}
\multiput(14.35,2.85)(1.5,0){3}{\vector(-1,-1){1.2}}
\multiput(3.85,1.35)(1.5,0){5}{\vector(-1,-1){1.2}}
\multiput(14.35,1.35)(1.5,0){3}{\vector(-1,-1){1.2}}


\end{picture} \\
    \vspace*{6ex}Table M
    \label{tab:M}
  \end{center}
\end{table}

\subsection{}
\label{sec:5.3}

The first hint that the Lie superalgebra $E(3|6)$ is somehow related to
the Standard  Model comes from the observation that its subalgebra
$\fa_0$ is isomorphic to the complexified Lie algebra of the group of
symmetries of the Standard Model.  Table~P below lists all
$\fa_0$-multiplets of fundamental particles of the Standard Model (see
e.g.~\cite{O}):  the upper part is comprised of three generations of
quarks and the middle part of three generations of leptons (these are
all fundamental fermions from which matter is built), and the lower part
is comprised  of the fundamental bosons (which mediate the strong and
electro-weak interactions).

  \begin{center}

\begin{tabular}{cc|ccc}
multiplets & charges  && particles\\
\hline \\[-1ex]
$(01,1,\,\,\tfrac{1}{3})$ & $\,\,\,\tfrac{2}{3},-\tfrac{1}{3}$ &
$\binom{u_L}{d_L}$ & $\binom{c_L}{s_L}$
    & $\binom{t_L}{b_L}$\\[1ex]
$(10,1,-\tfrac{1}{3})$ & $-\tfrac{2}{3},\,\tfrac{1}{3}$ &
$\binom{\tilde{u}_R}{\tilde{d}_R}$ &
     $\binom{\tilde{c}_R}{\tilde{s}_R}$ & $\binom{\tilde{t}_R}{\tilde{b}_R}$\\[1ex]
$(10,0,-\tfrac{4}{3})$ & $-\tfrac{2}{3}$ & $\tilde{u}_L$ &
$\tilde{c}_L$
    & $\tilde{t}_L$\\[1ex]
$(01,0,\,\,\tfrac{4}{3})$ & $\,\,\,\tfrac{2}{3}$ & $u_R$ & $c_R$ & $ t_R$\\[1ex]
$(01,0,-\tfrac{2}{3})$ & $-\tfrac{1}{3}$ & $d_R$ & $s_R$ &$b_R$\\[1ex]
$(10,0,\,\,\tfrac{2}{3})$ & $\,\,\,\tfrac{1}{3}$ & $\tilde{d}_L$ &
$\tilde{s}_L$ & $   \tilde{b}_L$\\[-1ex]
%
\setlength{\unitlength}{0.1in}
\begin{picture}(8.5,3)(0,-1)
  \multiput(0,0)(2,0){5}{\line(1,0){1.5}}
\end{picture}
& \setlength{\unitlength}{0.1in}
\begin{picture}(8.5,3)(0,-1)
  \multiput(0,0)(2,0){5}{\line(1,0){1.5}}
\end{picture}
& \setlength{\unitlength}{0.1in}
\begin{picture}(5,3)(1,-1)
  \multiput(0,0)(2,0){5}{\line(1,0){1.5}}
\end{picture}
& \setlength{\unitlength}{0.1in}
\begin{picture}(5,3)(0,-1)
  \multiput(0,0)(2,0){5}{\line(1,0){1.5}}
\end{picture}
& \setlength{\unitlength}{0.1in}
\begin{picture}(5,3)(0,-1)
  \multiput(0,0)(2,0){3}{\line(1,0){1.5}}
\end{picture}
\\
$(00,1,-1)$ & $0,-1$ & $\binom{\nu_L}{e_L}$
   & $\binom{\nu_{\mu L}}{\mu_L}$ & $\binom{\nu_{\tau
       L}}{\tau_L}$\\[1ex]
$(00,1,\,\,1)$ & $0,\,1$ & $\binom{\tilde{\nu}_R}{\tilde{e}_R}$
  & $\binom{\tilde{\nu}_{\mu R}}{\tilde{\mu}_R}$
  & $\binom{\tilde{\nu}_{\tau R}}{\tilde{\tau}_R}$\\[1ex]
$(00,0,\,\,2)$ & $\,\,\,1$ & $\tilde{e}_L$ & $\tilde{\mu}_L$
   & $\tilde{\tau}_L$\\[1ex]
$(00,0,-2)$ & $-1$ & $e_R$ & $\mu_R$ & $\tau_R$\\[1ex]
\hline \\[-1ex]
$(11,0,\,\,0)$ & $\,\,\,0$ & gluons\\[1ex]
$(00,2,\,\,0)$ & $1,-1,0$ & $W^+,W^-,Z$ & (gauge bosons)\\[1ex]
$(00,0,\,\,0)$ & $\,\,\,0$ & $\gamma$ & (photon)\\[1ex]
%
%
\end{tabular}
\begin{center}
  Table P
\end{center}

  \end{center}

It is easy to deduce from Theorem~4 that this list of multiplets
(plus the multiplets $(11,0,\pm 2)$) is characterized by the
conditions:

\romanparenlist
\begin{enumerate}
\item 
  $\fa_0$-multiplet occurs in a degenerate irreducible
  $E(3|6)$-module,
\item 
  when restricted to $s\ell_3 \subset \fa_0$, this multiplet
  contains only $1$-dimensional, the two fundamental or the
  adjoint representation,

\item 
$|Q| \leq 1$ for all particles of the multiplet, where the charge
$Q$ of a particle is given by the Gell-Mann-Nishijima formula:
$Q=\tfrac{1}{2} (y+h)$, where $y$ (resp.~$h$) is the
$Y$-eigenvalue (resp. $H=\diag (1,-1) \in s\ell_2$-eigenvalue).

\end{enumerate}

How can we see the number of generations of quarks and leptons?
For that order the sequence subcomplexes in Table~M by $t=r-q+3$
in sector~IV (time), and replace in them the induced modules by
their irreducible quotients.  Then we find \cite{KR2} (based on
computer calculations by Joris Van der Jeugt) that a fundamental
particle multiplet appears in the $t$\st{th} sequence iff $t \geq
1$.  Furthermore, for $1 \leq t \leq 7$ we get sequences with
various particle contents, but for $t \geq 8$ the particle
contents remains unchanged, and it is invariant under the CPT
symmetry (though for $t \leq 7$ it is not).  The explicit contents
is exhibited in \cite{KR2}, 659--660.

Remarkably, precisely three generations of leptons occur in the
stable region $(t \geq 8)$, but the situation with quarks is more
complicated:  this model predicts a complete fourth generation of
quarks and an incomplete fifth generation (with missing down type
triplets).

In view of this discussion, it is natural to suggest that the
algebra $su_3 +su_2 + u_1$ of internal symetries of the
Weinberg-Salam-Glashow Standard Model extends to $E(3|6)$.  It is
hoped that the representation theory of $E(3|6) $ will shed new
light on various features of the Standard Model.  I find it quite
remarkable that the $SU_5$ Grand Unified Model of Georgi-Glashow
combines the left multiplets of fundamental fermions in precisely
the negative part of the consistent gradation of $E(5|10)$ (see
Sec.~1.6).  This is perhaps an indication of the possibility that
the extension from $su_5$ to $E(5|10)$ algebra of internal
symmetries may resolve the difficulties with the proton decay.

\label{lastpage}

\end{document}